%% file: paper_arxiv.tex
\documentclass[12pt]{article}
\include{_preamble}

\zexternaldocument*[supp:]{sm}

\makeatletter
\@ifclassloaded{biometrika}{%
  \let\document\latexdocument
  \let\enddocument\latexenddocument
  \AtEndDocument{\printhistory}
  \let\arabic\latexarabic
  \def\rm{}

  \authorlist
}{%
  \author{\authorlist}
  \date{\today}
}
\makeatother
\title{\titlepaper}

\begin{document}
\maketitle

\begin{abstract}
  Continuous treatments have posed a significant challenge for causal inference,
  both in the formulation and identification of scientifically meaningful
  effects and in their robust estimation. Traditionally, focus has been placed
  on techniques applicable to binary or categorical treatments with few levels,
  allowing for the application of propensity score-based methodology with
  relative ease. Efforts to accommodate continuous treatments introduced the
  generalized propensity score, yet estimators of this nuisance parameter
  commonly utilize parametric regression strategies that sharply limit the
  robustness and efficiency of inverse probability weighted estimators of causal
  effect parameters. We formulate and investigate a novel, flexible estimator of
  the generalized propensity score based on a nonparametric function estimator
  that provably converges at a suitably fast rate to the target functional so as
  to facilitate statistical inference. With this estimator, we demonstrate the
  construction of nonparametric inverse probability weighted estimators of
  a class of causal effect estimands tailored to continuous treatments. To
  ensure the asymptotic efficiency of our proposed estimators, we outline
  several non-restrictive selection procedures for utilizing a sieve estimation
  framework to undersmooth estimators of the generalized propensity score. We
  provide the first characterization of such inverse probability weighted
  estimators achieving the nonparametric efficiency bound in a setting with
  continuous treatments, demonstrating this in numerical experiments.
  We further evaluate the higher-order efficiency of our proposed estimators by
  deriving and numerically examining the second-order remainder of the
  corresponding efficient influence function in the nonparametric model.
  %Using data from a recent vaccine efficacy trial, we illustrate the use of our
  %generalized propensity score and efficient estimators in ascertaining the
  %causal impacts of modulating post-vaccination immune responses on disease
  %risk.
  Open source software implementing our proposed estimation techniques, the
  \texttt{haldensify} \texttt{R} package, is briefly introduced.
\end{abstract}

%%%%%%%%%%%%%%%%%%%%%%%%%%%%%%%%%%%%%%%%%%%%%%%%%%%%%%%%%%%%%%%%%%%%%%%%%%%%%%%
\section{Introduction}\label{intro}

Across a range of scientific fields, research efforts often aim to quantify the
causal effects of intervening on continuous or ordinal treatments. Examples
include evaluating the impacts of increased physical exercise on
aging~\citep{diaz2012population}, reductions in surgical operating time on
post-surgical health outcomes~\citep{haneuse2013estimation}, changes in
vaccine-induced immunologic response on disease
risk~\citep{hejazi2020efficient}, impact of nurse hours per patient on hospital
readmission risk~\citep{mchugh2013hospitals}, and mortality reduction from
COVID-19 delay-in-intubation policies~\citep{diaz2022causal}. In these settings,
causal effects are often quantified in terms of \textit{causal} dose-response
curves~\citep{imbens2000role, kennedy2017nonparametric}. Unfortunately, the
estimation of such dose-response curves for continuous treatments in
infinite-dimensional statistical models is exceptionally challenging. Estimators
of these curves generally do not enjoy $n^{1/2}$-rate asymptotic behavior and
are incompatible with many standard estimation techniques.

To circumvent these challenges, it is common in practice to discretize the
treatment into categories, which enables practitioners to recover $n^{1/2}$-rate
behavior with standard estimation techniques. However, this strategy is subject
to a host of limitations, the most fundamental of which is a potential violation
of the causal identification assumption of consistency. This assumption
stipulates that hypothetical interventions should be well-defined, which is
rarely (if ever) the case when conceptualizing interventions that assign
a continuous treatment to a set of (discrete) values. More practically, if the
treatment is poorly discretized, perhaps due to limited subject matter
knowledge, the resultant dose-reponse curve is prone to providing a poor
approximation of the underlying curve.

More recent work has developed asymptotic theory under a range of semiparametric
assumptions. For example, \citet{kennedy2017nonparametric} proposed estimators
based on locally linear smoothing; \citet{vdl2018cvtmle} considered
cross-validated targeted minimum loss estimation of a general class of
non-standard parameters, including the dose-response curve as an example; and
\citet{westling2020causal} developed estimators assuming monotonicity of the
dose-response curve. These approaches represent valuable contributions in this
area. However, each approach relies on assumptions that are difficult to justify
across a range of settings. In particular, each posits a positivity assumption
requiring common support of the covariate-conditional treatment density across
all possible covariate values.

This fundamental limitation has motivated the study of causal effects of
\textit{stochastic} interventions~\citep{stock1989nonparametric,
diaz2012population, haneuse2013estimation, young2014identification}. These
interventions consider setting each individual's treatment level to a random
draw from an investigator-specified distribution.
%This approach makes for a highly
%flexible means of defining counterfactual random variables --- even static
%interventions are a special case (i.e., in which the post-intervention treatment
%value is drawn from a degenerate distribution with all mass placed on a single
%treatment level).
By carefully considering the candidate post-intervention distribution of the
treatment, practitioners can avoid making unrealistic positivity assumptions.
One popular approach draws treatment values from a modification of the naturally
occurring treatment distribution. Counterfactuals defined in this way are better
aligned with plausible future interventions that may be engineered by scientific
or policy initiatives. Recent efforts have yielded several candidate
approaches~\citep[e.g.,][]{diaz2012population, haneuse2013estimation,
young2014identification, kennedy2019nonparametric} for identifying and
estimating the causal effects of stochastic interventions, including for
longitudinal treatment regimens~\citep{diaz2021lmtp,diaz2022causal} and
mediation analysis~\citep[e.g.,][]{diaz2020causal, diaz2020nonparametric,
hejazi2020nonparametric}.

Many estimation approaches for stochastic interventions center around estimation
of the \textit{generalized} propensity score~\citep{hirano2004propensity}, the
conditional treatment density given covariates. Estimation of the generalized
propensity is a common problem in causal inference and has been studied in the
context of, for example, marginal structural models~\citep{robins2000marginal},
and covariate balancing~\citep{hirano2004propensity}. Despite its prevalence,
most proposals rely on assumptions imposing restrictive parametric forms to
facilitate estimation of this quantity, though flexible estimators leveraging
advances in machine learning have emerged~\citep[e.g.,][]{diaz2011super,
zhu2015boosting}. Numerical studies examining the impact of generalized
propensity score estimation on the evaluation of the causal effects of
continuous treatments were contributed by~\citet{austin2018assessing}.

In the present work, we develop a flexible, nonparametric estimator of the
generalized propensity score that is fully compatible with techniques for sieve
estimation. This estimator is formulated with the goal of evaluating the causal
effects of continuous treatments, using the framework of stochastic
interventions. Building upon this flexible nuisance estimator, we formulate and
evaluate a unique class of nonparametric inverse probability weighted (IPW)
estimators capable of asymptotically achieving a semiparametric analog of the
efficiency bound. This represents, to the best of our knowledge, two significant
theoretical and practical advances: (1) successfully tailoring undersmoothing
principles to conditional density functionals for the efficient estimation of
low-dimensional statistical parameters, and (2) developing IPW estimators of
continuous treatment effects that are both asymptotically efficient and
competitive with modern doubly robust estimators. We additionally discuss the
open source \texttt{haldensify}~\citep{hejazi2021haldensify} package, for the
\texttt{R} statistical programming language and environment~\citep{R}, which
implements our generalized propensity score and IPW estimators.

The remainder of the present manuscript is organized as follows.
Section~\ref{prelim} introduces notation and discusses existing theory on the
causal effects of stochastic interventions and their efficient estimation.
Section~\ref{methods} outlines our proposed conditional density estimation
procedure and goes on to detail its use in constructing a novel class of
nonparametric IPW estimators, proposing several procedures for sieve estimation
designed to allow these estimators to attain asymptotic efficiency.
Section~\ref{sim} investigates the relative performance of our proposed
nonparametric IPW estimators in numerical experiments.
%Section~\ref{application} demonstrates the application of our estimators in
%evaluating immune correlates of protection in the HVTN 505 trial.
Section~\ref{discuss} concludes with a discussion of future avenues of
investigation.

%%%%%%%%%%%%%%%%%%%%%%%%%%%%%%%%%%%%%%%%%%%%%%%%%%%%%%%%%%%%%%%%%%%%%%%%%%%%%%%
\section{Preliminaries}\label{prelim}

\subsection{Problem formulation and notation}\label{setup}

Let $W \in \mathcal{W}$ denote a vector of baseline covariates, $A \in
\mathcal{A}$ a real-valued continuous (or ordinal) treatment, and $Y \in
\mathcal{Y}$ an outcome of interest. To formalize the causal question of
interest, we introduce a structural causal model (SCM) to
describe the data-generating process~\citep{pearl2009causality}. Specifically,
we assume the following system of structural equations generates the observed
data for a single unit:
\begin{equation*}\label{npsem}
  W = f_W(U_W); A = f_A(W, U_A); Y = f_Y(A, W, U_Y),
\end{equation*}
where $\{f_W, f_A, f_Y\}$ are deterministic functions, and $\{U_W, U_A, U_Y\}$
are exogenous random variables. Importantly, the SCM implies a model for the
distribution of counterfactual random variables, which are generated by specific
interventions on the data-generating process. The standard identification
assumptions of consistency ($Y^{d(a, w)} = Y$ in the event $A = d(a, w)$) and
lack of interference~\citep{cox1958planning}
%($Y^{d(a_i, w_i)}_i \indep d(a_j, w_j)$ for $i \neq j$) \id{something odd here,
%the right hand side of the independence statement is not a random variable.
%Also, this states that treatment of unit $j$ does not affect outcomes of unit
%$i$, but do you not need other types of no-interference (e.g., outcomes on
%outcomes). I would suggest to delete this and just assume we are samping from
%an iid NPSEM, which implies no interference.}
hold as derived properties of this SCM~\citep[for a discussion on the former,
see][]{pearl2010brief}. The complete (unobserved) data
unit~\citep{neyman1938contribution} is $X = (W, Y_a: a \in \mathcal{A})$, where
the counterfactuals $Y_a$ are the potential outcomes corresponding to the values
the outcome would take under each possible value in the support of the treatment
$\mathcal{A}$. We will focus on the estimation of counterfactual treatment
parameters that are functionals of the distribution of $X$.

We consider the data on a single observational unit $O$, denoted $O = (W, A,
Y)$, as arising from $P_0$, the true generating distribution of $O$.
Assuming access to $n$ i.i.d.~copies of $O$, we use $P_n$ to denote the
empirical distribution of $O_1, \ldots, O_n$. At times, it will prove convenient
to use standard notation from empirical process theory; specifically, we will
write $P_n f \coloneqq n^{-1} \sum_{i = 1}^n f(O_i)$ and, more generally,
$Pf \coloneqq \int f(O) dP$ on occasion. Assuming only that $P_0$ is an element
of the nonparametric statistical model $\mathcal{M}$, i.e., $P_0 \in
\mathcal{M}$, we strive to avoid placing undue restrictions on the form of
$P_0$.
We use $p_0$ to denote the density of $O$, which evaluated on a typical
observation $o$, is
\begin{equation*}\label{likelihood_factorization}
  p_0(o) = q_{0,Y}(y \mid A = a, W = w) g_{0,A}(a \mid W = w) q_{0,W}(w) \ ,
\end{equation*}
where $q_{0, Y}$ denotes the conditional density of $Y$ given $\{A, W\}$ with
respect to some dominating measure, $g_{0, A}$ the conditional density of $A$
given $W$ with respect to dominating measure $\mu$, and $q_{0, W}$ the density
of $W$ with respect to dominating measure $\nu$.

Counterfactual quantities of interest may be defined by specific interventions
that alter the structural equation $f_A$ and ``surgically'' insert
post-intervention treatment values in place of those naturally generated by
$f_A$. \textit{Static interventions} make for a familiar example: $f_A$ is
merely replaced with a specific value, selected \textit{a priori}, $a \in
\mathcal{A}$. When $\mid\mathcal{A}\mid$, the cardinality of $\mathcal{A}$, is
small, contrasts of the counterfactual means of static interventions under each
$a \in \mathcal{A}$ can prove informative. On the other hand, when
$\mid\mathcal{A}\mid$ is large or uncountable (e.g., when $A \in \R$, i.e., is
continuous), the evaluation of many such counterfactuals is both of questionable
scientific relevance and is theoretically challenging.

A \textit{stochastic intervention} modifies the value $A$ would naturally
assume, replacing it with a draw from a post-intervention distribution
$\tilde{g}_{0,A}(\cdot \mid W)$ (n.b., the naught subscript emphasizes that this
distribution may depend on the true, but unknown, data-generating distribution
$P_0$). A stochastic intervention may be designed to collapse into a static
intervention by selecting $\tilde{g}_{0,A}(\cdot \mid W)$ to be degenerate,
placing all mass on a single point $a \in \mathcal{A}$. The taxonomy of
stochastic interventions is, as of yet, unsettled --- the term is widely used to
describe intervention regimes exhibiting randomness conditional on covariates
$W$. From this perspective, static interventions are deterministic and dynamic
interventions, which assign treatment based on $W$, are deterministic
conditional on $W$. Throughout, we focus on intervention regimes meeting this
latter definition.

Using the former definition,~\citet{diaz2012population} described a stochastic
intervention that draws $A$ from a distribution such that $\tilde{g}_{0,A}(a
\mid W) = g_{0,A}(a - \delta(W) \mid W)$ for all $a \in \mathcal{A}$.
%\id{
%We still need to tidy up this notation. For example, the function in
%(\ref{eqn:shift_additive}) has no inverse. I recommend that we use the notation
%I used in the lmtp paper.
%}
%\nh{
%I've modified the claim about the 2012 paper to use the exact definition given
%there (removing the inverse of the shifting function), and I've modified some
%of the discussion below to better position the contributions of Rotnitzky and
%Haneuse. Also, unless i'm missing something critical, then i'd disagree with
%the claim that \eqref{eqn:shift_additive} isn't invertible --- in fact, it's
%trivially invertible since each ``piece'' is itself invertible.
%}
Shortly thereafter,~\citet{haneuse2013estimation}, taking the latter definition,
showed that estimation of the causal effect attributable to the stochastic
intervention of~\citet{diaz2012population} is equivalent to the effect of an
investigator-supplied hypothetical intervention modifying the value $A$ would
naturally assume according to a regime $d(a,w;\delta) = a + \delta(w)$. These
authors go on to characterize a broad class of interventions by introducing the
assumption of \textit{piecewise smooth invertibility}.
\begin{assumption}[Piecewise smooth invertibility]\label{ass:piece_inv}
  For each $w \in \cal W$, assume that the interval
  ${\cal I}(w) = (l(w,), u(w))$ may be partitioned into subintervals
  ${\cal I}_{\delta,j}(w):j = 1, \ldots, J(w)$ such that the regime
  $d(a, w; \delta)$ is equal to some $d_j(a, w; \delta)$ in ${\cal
  I}_{\delta,j}(w)$ and $d_j(\cdot, w; \delta)$ has inverse function
  $b_j(\cdot, w; \delta)$ with derivative $b_{j}'(\cdot, w; \delta)$.
\end{assumption}
Assumption~\ref{ass:piece_inv} can be used to show that the intervention implied
by the regime $d(A,W;\delta)$ may be interpreted as acting on the individual
level~\citep{young2014identification}, incompatible with the definition used
by~\citet{diaz2012population}. Importantly, the regime $d(A,W;\delta)$ may
depend on both covariates $W$ and the treatment $A$ that would be assigned in
its absence (i.e., the treatment's \textit{natural} value); consequently, such
regimes have been termed \textit{modified treatment policies} (MTP). Regimes
obeying the MTP definition constitute an important class of stochastic
interventions of the second variety (n.b., these are non-deterministic
conditional on $W$, since they depend on $A$).
Both~\citet{haneuse2013estimation} and~\citet{diaz2018stochastic} were motivated
by counterfactual questions of modifying an assigned treatment, the former
seeking to evaluate the effect on patient health of reducing surgical operating
time and the latter by the effect of adjusting prescribed exercise regimens
based on existing athletic habits. Conveniently, both sets of authors considered
an MTP of the form:
\begin{equation}\label{eqn:shift_additive}
  d(a, w; \delta) =
  \begin{cases}
    a + \delta(w) & \text{if } a + \gamma \leq u(w) \\
    a & \text{if } a + \gamma > u(w)
  \end{cases},
  \text{where} \quad \delta(w) = \gamma \in \R
\end{equation}
and $u(w)$ is the maximum value in the conditional support of $g_{0,A}(\cdot
\mid W = w)$. While an intuitively useful MTP, in many settings, an additive
alteration to the natural value of $A$ may prove inadequate to formulate an
actionable policy. This arises, for example, in environmental health, in which
it is of interest to evaluate effects of multiplicative reductions to, say,
pollutants. MTPs can be tailored to such uses-cases easily:
\begin{equation}\label{eqn:shift_mult}
  d(a, w; \delta) =
  \begin{cases}
    a \times \delta(w) & \text{if } a \times \gamma \leq u(w) \\
    a & \text{if } a \times \gamma > u(w)
  \end{cases},
  \text{where} \quad \delta(w) = \gamma \in \R \ .
\end{equation}
The MTPs in equations~\eqref{eqn:shift_additive} and~\eqref{eqn:shift_mult} can
be seen as generating a counterfactual outcome
$Y_{d(A,W; \delta)} \coloneqq f_Y(d(A, W; \delta), W, U_Y)$ whose distribution
we denote $P_0^{\delta}$. The goal, then, is to identify and estimate $\psi_{0,
\delta} \coloneqq \E_{P_0^{\delta}} \{Y_{d(A,W; \delta)}\}$, the mean of this
counterfactual outcome.
%\id{here you can say that the formula for the density $\tilde g$ as a function
%of $g$ for an MTP satisfying A1 is given in \cite{haneuse2013estimation}}
%\nh{resolved below by citing HR and the LiDA paper}

\subsection{Identification of the population intervention
  effect}\label{pie_param}

Using the population intervention effect (PIE)
framework~\citep{hubbard2008population}, \citet{diaz2012population,
diaz2018stochastic} introduced the PIE for stochastic interventions $\theta_{0,
\delta} \coloneqq \psi_{0, \delta} - \E Y$. As $\E Y$ is trivially estimable,
their efforts focused on identifying and estimating $\psi_{0,\delta}$. For the
MTP~\eqref{eqn:shift_additive}, these authors showed $\psi_{0, \delta}$ to be
identified by
\begin{align}\label{eqn:identification2012}
  \psi_{0,\delta} &= \int_{\mathcal{W}} \int_{\mathcal{A}}
  \overline{Q}_{0,Y}(a, w) \tilde{g}_{0, A}(a \mid W = w)
  q_{0, W}(w) d\mu(a)d\nu(w) \nonumber \\
  &= \int_{\mathcal{W}} \int_{\mathcal{A}}
  \overline{Q}_{0,Y}^{d}(a, w) g_{0, A}(a \mid W = w)
  q_{0, W}(w) d\mu(a)d\nu(w),
\end{align}
%\id{here and throughout we are going to have to say that the above is an abuse
%of notation (i.e., using $d^{-1}$), as the above (and the density ratios
%defining H below) is not exactly correct under A1}
%\nh{is your point here that A1 does not guarantee that the counterfactual
%treatment distribution $\tilde{g}(A, W)$ may not be equivalent to
%$g(d^{-1}(a, w), W)$? I've changed the above and made a note about the abuse
%of notation that follows.}
where $\overline{Q}_{0,Y}(a,w) \coloneqq \E_{P_0} \{Y \mid A = a, W = w\}$
(conditional mean of $Y$ given $A = a$ and $W = w$, as implied by $P_0$),
$\overline{Q}_{0,Y}^{d}(a,w) \coloneqq \overline{Q}_{0,Y}(d(a, w; \delta), w)$,
and $g_{0,A}(a \mid W = w)$ is the conditional density of the treatment. We
stress that the equivalence in equation~\eqref{eqn:identification2012} holds
only when the post-intervention treatment density $\tilde{g}_{0, A}$ arises from
a regime $d(A,W;\delta)$ adhering to Assumption~\ref{ass:piece_inv}, which
facilitates use of the change-of-variable formula in the integral expressions.
This mathematical convenience allows identification of $\psi_{0,\delta}$ by
replacing evaluation of the expectation $\overline{Q}_{0,Y}(a,w)$ under the
counterfactual treatment density $\tilde{g}_{0, A}$ with evaluation of the
post-modification outcome mechanism $\overline{Q}_{0,Y}^{d}(a,w)$ under the
natural treatment density $g_{0,A}$. This has proven to be a significant
simplification, essential for the formulation of efficient estimators that are
computationally feasible~\citep[e.g.,][]{diaz2018stochastic,
hejazi2020efficient}. Considerations necessary for appropriately expressing
$\tilde{g}_{0, A}$ in terms of $g_{0,A}$ for the MTPs~\eqref{eqn:shift_additive}
and~\eqref{eqn:shift_mult}, whether $A$ is continuous or discrete-valued, have
been previously outlined by~\citet{haneuse2013estimation} and~\citet[][see
equations (6) and (7)]{diaz2022causal}. Throughout, in a minor abuse of
notation, we will use $\tilde{g}_{0, A}$ to refer to the appropriate
post-intervention treatment density for the MTPs~\eqref{eqn:shift_additive}
and~\eqref{eqn:shift_mult}.

For the statistical functional~\eqref{eqn:identification2012} to correspond to
the causal estimand $\psi_{0,\delta}$, several standard but untestable
assumptions are required, including the following.
% Identification assumptions
%\begin{assumption}[Lack of interference]
  %Assume $Y^{d(a_i, w_i)}_i \indep d(a_j, w_j)$ for $i \neq j$.
  %\label{ass:interfere}
%\end{assumption}
%\begin{assumption}[Consistency]
  %Assume $Y^{d(a, w)} = Y$ in the event $A = d(a, w)$.
  %\label{ass:consist}
%\end{assumption}
\begin{assumption}[No unmeasured confounding]
  Assume $A \indep Y^{d(a, w; \delta)} \mid W = w$.
  \label{ass:nuc}
\end{assumption}
\begin{assumption}[Positivity]
  Assume $a \in \mathcal{A} \implies d(a, w; \delta) \in \mathcal{A} \mid
  W = w~\forall~w \in \mathcal{W}$.
  \label{ass:pos}
\end{assumption}
%\id{The no interference and consistency assumptions are implied, since we are
%working with an i.i.d.~NPSEM.}
%Together, assumptions~\ref{ass:interfere} and~\ref{ass:consist} are often
%referred to as the stable unit treatment value assumption
%(SUTVA)~\citep{rubin1978bayesian,rubin1980randomization}.
The positivity assumption~\ref{ass:pos}, required to establish
equation~\eqref{eqn:identification2012}, is unlike its analog for simpler (i.e.,
static or dynamic) intervention regimes. Instead of requiring positive mass to
be placed across all treatment levels for all covariate strata $w \in
\mathcal{W}$, this positivity condition requires only that the
post-intervention treatment mechanism be bounded, i.e., $\prob_{P_0}
\{\tilde{g}_{0,A}(A \mid W) / g_{0,A}(A \mid W) < \infty \} = 1$, which may
generally be satisfied by a suitable choice of the parameter $\delta(W)$ in an
MTP $d(A,W; \delta)$. Importantly, this assumption ensures that the
post-intervention treatment density $\tilde{g}_{0,A}$ does not place mass at
points in the support at which $g_{0,A}$ places no mass, leading to well-defined
counterfactuals.
%\id{That would translate into the ratio being bounded instead of it being
%positive.}

\subsection{Efficient influence function of the population intervention
effect}\label{pie_eif}

Beyond their careful study of the identification of this causal effect,
\citet{diaz2012population, diaz2018stochastic} derived the \textit{efficient
influence function} (EIF), a quantity central to semiparametric efficiency
theory, of $\psi_{0, \delta}$ in the nonparametric model $\M$. Using the EIF,
these authors proposed asymptotically efficient doubly robust estimators
constructed based on the form of the EIF. When evaluated on a typical
observation $o$, the EIF is
\begin{equation}\label{eqn:eif_full}
  D^{\star}(o; P_0) = H_0(a, w)\{y - \overline{Q}_{0,Y}(a, w)\} +
  \overline{Q}_{0,Y}^{d}(a, w) - \psi_{0,\delta},
\end{equation}
where the auxiliary covariate $H(a,w)$ takes the form $H_0(a, w) =
\tilde{g}_{0, A}(a \mid w) / g_{0, A}(a \mid w)$. The EIF characterizes the best
possible asymptotic variance, or nonparametric efficiency bound, of all regular
asymptotically linear estimators of $\psi_{0, \delta}$ and is thus usually
a critical ingredient in the development of efficient estimation strategies.
While the expression of the EIF appearing in equation~\eqref{eqn:eif_full} is
suitable for the construction of efficient estimators, it arises from a von
Mises expansion~\citep{mises1947asymptotic, bickel1993efficient, vdl2003unified}
of the parameter functional $\Psi_{\delta}: \M \to \R$ at the distributions $P,
P_0 \in \M$:
\begin{align}\label{eqn:r2_expansion}
  n^{1/2}(&\Psi_{\delta}(P) - \Psi_{\delta}(P_0)) =\\~&n^{-1/2}
    P(D^{\star}(P_0) - D^{\star}(P)) +
    n^{1/2}(P - P_0)(D^{\star}(P) - D^{\star}(P_0)) \nonumber -
    n^{1/2} R_2(P, P_0) \ .
  %-P_0 D^{\star}(P) + R_2(P, P_0),
\end{align}
Equation~\eqref{eqn:r2_expansion} is critically important to characterizing the
asymptotic behavior of regular and asymptotically linear estimators of
$\Psi_{\delta}(P_0) \equiv \psi_{0,\delta}$. The first term can be shown to
converge readily to a zero-centered Gaussian distribution with variance equal to
the variance of the EIF, i.e., $N(0, P_0 D^{\star}(P_0)(O)^2)$, by the central
limit theorem. The second term, which corresponds to ``plug-in bias'', generally
fails to vanish asymptotically, motivating the development of specialized
correction strategies for this express purpose (e.g., the targeted learning
framework of~\citet{vdl2011targeted}). Under standard empirical
process conditions (on nuisance parameter estimators), the third and fourth
terms can be shown to converge to zero in probability; moreover,
cross-validation can be used to circumvent even these
conditions~\citep{klaassen1987consistent, zheng2011cross}. The last term
$R_2(P, P_0)$ of the expansion~\eqref{eqn:r2_expansion} is a second-order
remainder associated to the scaled difference of $\Psi_{\delta}(P)$ and
$\Psi_{\delta}(P_0)$; furthermore, it is typical to assume necessary rates of
nuisance parameter convergence such that $R_2(P, P_0) = o_p(n^{-1/2})$. Such an
assumption would be satisfied, for example, if all nuisance estimators were to
converge to their target functionals at $n^{-1/4}$--rate. Under standard
conditions, the asymptotic efficiency of regular asymptotically linear
estimators may be described via the EIF:
\begin{equation*}
  \psi_{n,\delta} - \psi_{0,\delta} = n^{-1}
    \sum_{i = 1}^n D^{\star}(P_0)(O_i) + o_p(n^{-1/2}),
\end{equation*}
where $\psi_{n,\delta} \equiv \Psi_{\delta}(P_n)$ is an asymptotically linear
estimator of the target parameter $\psi_{0,\delta}$. The form of the expansion
emphasizes that the EIF measures the quality of the estimator $\psi_{n,\delta}$,
and that $\psi_{n,\delta}$ must solve the EIF estimating equation to be
asymptotically efficient.

\subsection{Estimation of the population intervention effect}\label{pie_est}

For estimation of $\psi_{0,\delta}$, \citet{diaz2018stochastic} defined a direct
(or, substitution) estimator based on the G-computation formula. This estimator
is
\begin{align}\label{eqn:plugin}
  \psi_{n,\delta}^{\text{SUB}} \coloneqq&
      \int \overline{Q}_{n,Y}^{d}(a, w) dQ_{n,AW}(a,w) \nonumber \\
      =&~n^{-1} \sum_{i=1}^n \overline{Q}_{n,Y}^{d}(A_i, W_i) \ ,
\end{align}
where $Q_{n,AW}(a,w)$ is merely an appropriate estimate of the joint
distribution of $(A,W)$. An inverse probability weighted (IPW) estimator of
$\psi_{0,\delta}$ takes the form
\begin{align}\label{eqn:ipw}
  \psi_{n,\delta}^{\text{IPW}} &= n^{-1} \sum_{i=1}^n H_n(A_i, W_i) Y_i \\
    &= n^{-1} \sum_{i=1}^n \frac{\tilde{g}_{n, A}(A_i
    \mid W_i)}{g_{n, A}(A_i \mid W_i)} Y_i \nonumber \ .
\end{align}
Popular in practice, the stabilized IPW estimator, an adaptation of the
estimator of equation~\eqref{eqn:ipw}, uses standardization to reduce the impact
of unstable inverse weights:
\begin{equation}\label{eqn:ipw_stable}
  %\psi_{n,\delta}^{\text{IPW}} = n^{-1} \sum_{i=1}^n
  %\frac{\{\tilde{g}_{n, A}(A_i \mid W_i) / g_{n, A}(A_i \mid W_i)\}}{
  %n^{-1} \sum_{i=1}^n\{\tilde{g}_{n, A}(A_i \mid W_i) /
  %g_{n, A}(A_i \mid W_i)\}} Y_i .
  \psi_{n,\delta}^{\text{IPW}} = n^{-1} \sum_{i=1}^n
    \frac{H_n(A_i, W_i)}{\overline{H}_n(A, W)} Y_i,
\end{equation}
where $\overline{H}_n(A, W) = n^{-1} \sum_{i=1}^n H_n(A_i, W_i)$. In
equations~\eqref{eqn:plugin},~\eqref{eqn:ipw},~\eqref{eqn:ipw_stable}, and in
the sequel, the subscript $n$ denotes quantities estimated based on
the empirical distribution $P_n$ in lieu of their true counterparts (at $P_0
\in \M)$ --- that is, $g_{n,A}$ is an estimate of $g_{0,A}$, while
$\overline{Q}_{n,Y}$ is an estimate of $\overline{Q}_{0,Y}$. Occasionally,
subscripts are omitted entirely, in which case a given quantity should be taken
as being estimated at an arbitrary distribution $P \in \M$.

Relevant to our discussion of $H(A,W)$,~\citet{diaz2021lmtp}, attempting to
avoid the laborious task of density estimation, proposed direct estimation of
this ratio of inverse probability weights, utilizing a re-parameterization based
on the odds of a variable in an artificial dataset constructed so as to include
records for all units under both $\tilde{g}_{0,A}$ and $g_{0,A}$. Their approach
adapted earlier ideas exploited by, among others,~\citet{qin1998inferences},
\citet{cheng2004semiparametric}, and~\citet{vdl2018online} in the context of
causal inference and explored independently in the context of general machine
learning~\citep[e.g.,][]{sugiyama2010conditional, sugiyama2012density}.
Recently,~\citet[][see page 8]{nugent2021evaluating}, in demonstrating causal
inference with MTPs, praised the computational convenience of the direct ratio
estimation approach, relative to direct estimation of the conditional density
$g_{0,A}$. Yet, this approach is only applicable in \textit{limited settings}.
To see this, consider the following shortcomings of this approach. Firstly,
direct ratio estimation is prone to producing biased estimated weights, which
arise from the direct estimate taking nonzero values even when $\tilde{g}_{0,A}$
is truly zero. This amounts to an artificial (that is, algorithmically induced)
violation of the positivity assumption~\ref{ass:pos}, requiring manual
correction even in standard software implementations. Secondly, there are
numerous settings in which such a re-parameterization categorically fails,
including in the estimation of weights for stochastic interventions that are not
MTPs or in causal mediation analysis with and without
MTPs~\citep[e.g.,][]{diaz2020causal}. Thirdly, and perhaps most critically, the
ratio of inverse probability weights is not amenable to likelihood-based
estimation or inference, as it is not itself a component of the underlying
likelihood of the data-generating process. Lastly, such a representation leads
to ``artificial'' score terms that are incompatible with scores based on
$g_{0,A}$, severely complicating any efforts to derive semiparametric-efficient
estimators within several well-studied frameworks.

%\mvdl{
%In particular, proving that an IPW estimator would be asymptotically efficient
%with HAL as shown here and in other paper would be problematic and likely not
%true. We might do a little explanation here. This method creates artificially
%two samples, $\Delta, A,W, \Delta=1$ then $A$ was drawn from $g^{\star}$ and if
%$\Delta=0$ then $A$ was drawn from $g$. Given this sample we fit
%$\Pr(\Delta=1 \mid A,W)$ with, say, HAL. Then $g^{\star}/g = \Pr(\Delta=1
%\mid A,W) / \Pr(\Delta=0 \mid A,W) \Pr(\Delta=0) / \Pr(\Delta=1)$, so, if the
%two samples are balanced, then $g^{\star}/g = \Pr(\Delta=1 \mid A,W)
%/ \Pr(\Delta=0 \mid A,W)$. This HAL of $\Pr(\Delta=1 \mid A,W)$ solves score
%equations $P_n h(A,W)(\Delta - \Pr(\Delta=1 \mid A,W))$; however, for making
%the IPTW estimator efficient we need to solve scores $P_n S(A,W) - \E_g(S \mid
%W)=0$. There is no way to write these real $g$ scores in terms of these scores
%of $S(\Delta,A,W)$. Another issue is that by taking equal size sample from
%$g^{\star}$ we create artificial noise since we know $g^{\star}$, so, clearly,
%it cannot be efficient. You might then say, lets the sample of $g^{\star}$
%dominate so that $\Pr(\Delta=1) \approx 1$ but then we are fitting
%$\Pr(\Delta=1 \mid A,W)$ with a rare outcome (or equivalently always $1$
%outcome), which is terribly noisy. So, we see there is no way this procedure can
%be efficient and likelihood-based.
%}

Both the direct estimator $\psi_{n,\delta}^{\text{SUB}}$ and the IPW estimator
$\psi_{n,\delta}^{\text{IPW}}$ require estimation of only a single nuisance
parameter --- the outcome mechanism and the conditional treatment density,
respectively --- but are well-known to be irregular when the corresponding
nuisance parameter is estimated flexibly, suffer from residual asymptotic bias,
and generally fail to achieve the nonparametric efficiency bound. What's more,
neither estimator is asymptotically linear for $\psi_{0,\delta}$ when flexible
regression strategies are used for nuisance parameter estimation, significantly
limiting the real-world scenarios in which these approaches may be successfully
applied. These notable issues, among others, have contributed to the rise in
popularity of doubly robust estimators; accordingly, we briefly review two such
estimators of $\psi_{0,\delta}$ in Section~\ref{supp:dr_est} of the
\href{sm}{Supplementary Materials}.
We focus instead on a nonparametric estimator $g_{n,A}$ of the generalized
propensity score $g_{0,A}$, which facilitates construction of nonparametric,
asymptotically efficient IPW estimators of $\psi_{0, \delta}$, when coupled with
sieve estimation and semiparametric efficiency theory.

\subsection{The highly adaptive lasso estimator}\label{hal}

In our subsequent developments, we make use of a nonparametric function
estimator, the highly adaptive lasso (HAL)~\citep{vdl2017generally,
vdl2017uniform}, which approximates a functional parameter of interest (e.g.,
conditional mean, hazard, density, conditional density) using a linear
combination of basis functions. HAL requires that the target functional belong
to the set of c\`{a}dl\`{a}g (i.e., right-hand continuous with left-hand limits)
functions with sectional variation norm bounded by a finite (but unknown)
constant, a \textit{global} constraint that limits the degree of variability
that the target functional can exhibit. Similar nonparametric estimation
approaches generally make use of \textit{local} smoothness assumptions. While
developments similar to ours may be possible under alternative assumptions,
pursuing such a course of study lies outside our scope here.

For any function $f \in \D[0,\tau]$, the Banach space of $d$-variate real-valued
c\`{a}dl\`{a}g functions on a cube $[0,\tau] \in \R^d$, the sectional variation
norm of $f$ may be expressed
\begin{equation*}
  \lVert f \rVert^{\star}_\nu \coloneqq \lvert f(0) \rvert + \sum_{s
  \subset\{1, \ldots, d\}} \int_{0_s}^{\tau_s} \lvert df_s(u_s) \rvert,
\end{equation*}
where $s$ are subsets of $\{0, 1, \ldots, d\}$, defined by partitioning
$[0,\tau]$ in $\{0\} \{ \cup_s (0_s,\tau_s]\}$, with the sum taken over all
subsets of the coordinates $\{0,1,\ldots,d\}$. For a given subset $s \subset
\{0,1,\ldots,d\}$, define $u_s \coloneqq (u_j : j \in s)$ and $u_{-s} \coloneqq
(u_j : j \notin s)$; then, $f_s: [0_s, \tau_s] \rightarrow \R$, defined as
$f_s(u_s) \coloneqq f(u_s,0_{-s})$. Thus, $f_s(u_s)$ is a section of $f$ that
sets the components not in $s$ to zero, that is, allowing variation only along
those components belonging to $u_s$. Per~\citet{qiu2021universal}, this
definition of variation norm corresponds with the notion of Hardy-Krause
variation~\citep{owen2005multidimensional}.

\citet{vdl2015generally, vdl2017generally} proved that the HAL estimator
converges to the target functional at a rate faster than $n^{-1/4}$, without any
contribution from the dimensionality $d$ of the problem at hand, so long as $d$
remains fixed. Subsequent theoretical investigations improved this rate of
convergence to $n^{-1/3} \log(n)^{d/2}$~\citep{bibaut2019fast}, with ongoing
work yielding promising further improvements still~\citep{vdl2022further}. The
HAL estimator proceeds in two steps. To start, a rich set of indicator (or
higher-order, i.e., spline) basis functions are generated to represent the
target functional; this step is a mapping of the covariate space in terms of the
HAL basis. Lasso regression~\citep{tibshirani1996regression} is then applied to
a linear combination of these basis functions, minimizing the expected value of
an appropriate loss function while constraining the $L_1$-norm of the vector of
coefficients to be bounded by a finite constant, its approximate sectional
variation norm). \citet{benkeser2016highly} demonstrated the utility of the HAL
estimator in an extensive series of numerical experiments. The HAL estimator is
implemented in the free and open source \texttt{hal9001}
package~\citep{coyle2021hal9001-rpkg, hejazi2020hal9001-joss} for the \texttt{R}
language and environment for statistical computing~\citep{R}.

When a nuisance parameter of interest is taken as the target functional, the HAL
estimator may be applied to generate initial estimates, under the assumption
that the true nuisance parameter functional (e.g., the generalized propensity
score) is of finite sectional variation. When the nuisance parameter $\eta$,
with arbitrary input $Z$, is real-valued, its HAL representation may be
expressed
\begin{align}\label{eq:hal}
  \eta(z) &= \eta(0) + \sum_{s \subset\{1,\ldots,d\}} \int_{0_s}^{z_s} d
  \eta_s(u_s) \nonumber \\ & = \eta(0) + \sum_{s \subset\{1,\ldots,d\}}
  \int_{0_s}^{\tau_s} \mathbb{I}(u_s \leq z_s) d \eta_s(u_s),
\end{align}
which can be approximated by the use of a discrete measure placing mass on each
observed $Z_{s,i}$. When the range of $\eta$ is the unit interval, an analogous
approach, using instead $\logit(\eta)$, may be leveraged based on
a representation of~\citet{gill1995inefficient}; notably,
\citet{ertefaie2020nonparametric} work with this representation in using HAL
regression for estimation of the propensity score in the case of binary
treatments. Now, take $Z_{s,i}$ to be support points of $\eta_s$ and let
$\phi_{s,i}(c_s) \coloneqq \mathbb{I}(Z_{s,i} \leq c_s)$, then
\begin{equation*}
 \eta_\beta = \beta_0 + \sum_{s \subset\{1,\ldots,d\}} \sum_{i=1}^{n}
   \beta_{s,i} \phi_{s,i},
\end{equation*}
where $\lvert \beta_0 \rvert + \sum_{s \subset\{1,\ldots,d\}} \sum_{i=1}^{n}
\lvert \beta_{s,i} \rvert$ is an approximation of the sectional variation norm
of $\eta$. The loss-based HAL estimator $\beta_n$ may then be defined
\begin{equation*}
  \beta_{n, \lambda} = \argmin_{\beta: \lvert \beta_0 \rvert + \sum_{s
  \subset\{1,\ldots,d\}} \sum_{i=1}^{n} \lvert \beta_{s,i} \rvert <\lambda}
  n^{-1} \sum_{i=1}^n L(\eta_{\beta})(O_i),
\end{equation*}
%\db{(Inconsistent notation. Above you write out sums when defining one-step.
%Pick one and use consistently.)}
where
%$P_n f = n^{-1} \sum_{i = 1}^n f(O_i)$ and
$L(\cdot)$ is an appropriately chosen loss function;
see~\citet{dudoit2005asymptotics} for an illuminating discussion of appropriate
choices of loss function for a range of loss-based estimation problems. Salient
to our goal of estimating the generalized propensity score, the negative
log-density loss, $L(\eta) = -\log(p_{\eta})$, is suitable for density
estimation. Finally, the HAL estimate of $\eta$ may be denoted
$\eta_{n,\lambda} \equiv \eta_{\beta_{n, \lambda}}$. Each choice of the
regularization term $\lambda$ corresponds to a unique HAL estimator, though
methods for the selection of $\lambda$ must be tailored to the estimation goal
in order to yield suitable candidate estimators.

%%%%%%%%%%%%%%%%%%%%%%%%%%%%%%%%%%%%%%%%%%%%%%%%%%%%%%%%%%%%%%%%%%%%%%%%%%%%%%%
\section{Efficient estimation with the generalized propensity score}\label{methods}

We now turn to procedures for estimation of the generalized propensity score
$g_{0,A}$, from which we may be able to subsequently develop efficient,
nonparametric estimators of $\psi_{0,\delta}$. For compatibility with our
subsequent theoretical results, we present our conditional density estimation
procedure in the context of the HAL estimator discussed in Section~\ref{hal};
this is chiefly for three reasons. Firstly, formal theoretical results guarantee
a suitable convergence rate for estimator construction when the HAL estimator is
used to approximate nuisance functionals~\citep{vdl2017generally}. Secondly,
contemporaneous efforts have made progress in developing efficient direct and
IPW estimators of low-dimensional parameters in causal inference
settings~\citep[e.g.,][]{vdl2019efficient,ertefaie2020nonparametric}. Thirdly,
the HAL regression algorithm is readily available and accessible via the free
and open source \texttt{hal9001} \texttt{R}
package~\citep{coyle2021hal9001-rpkg, hejazi2020hal9001-joss}.

In considering estimation of $g_{0,A}$, a straightforward strategy involves
assuming a parametric working model for a few moments of the density functional,
allowing the use of standard regression techniques to generate suitable
estimates~\citep{robins2000marginal, hirano2004propensity}. For example, one
could operate under the working assumption that the density of $A$ given $W$
follows a Gaussian distribution with homoscedastic variance and mean
$\sum_{j=1}^p \beta_j \phi_j(W)$, where $\phi = (\phi_j : j)$ are user-specified
basis functions and $\beta = (\beta_j : j)$ are unknown regression parameters.
Under such a regime, a density estimate could be generated by fitting a linear
regression of $A$ on $\phi(W)$ to estimate $\E[A \mid W]$, paired with maximum
likelihood estimation of the variance of $A$. In this case, the estimated
conditional density would be given by the density of a Gaussian distribution
evaluated at these estimates. Since such strategies make strong assumptions
about the form of the conditional density functional, they are more prone to
model misspecification than alternatives that strive to limit reliance on
structural assumptions.

Constructing a flexible density estimator is a more challenging problem, as the
set of available tools is limited. These limitations motivated our investigation
of conditional density estimation approaches capable of handling arbitrary
regression functions, which have been formulated and applied only sporadically
in causal inference~\citep[e.g.,][]{vdl2010targeted, diaz2011super,
sofrygin2019targeted}. We next describe a flexible class of nonparametric
estimators, implemented in the \texttt{haldensify} \texttt{R}
package~\citep{hejazi2021haldensify}; two more restrictive classes of
semiparametric conditional density estimators are outlined in
Section~\ref{supp:alt_dens} of the \href{sm}{Supplementary Materials}.

\subsection{Nonparametric estimator based on hazard
regression}\label{pooled_haz_est}

Estimators that eschew any assumptions on the form of the conditional density
are a rarity. Notably,~\citet{vdl2010targeted} and~\citet{diaz2011super}
proposed constructing estimators of this nuisance functional by a combination of
pooled regression and exploitation of the relationship between the (conditional)
hazard and density functions. Our proposal builds upon theirs, replacing their
recommendation of an arbitrary binary regression model with HAL regression. This
requires the key change of adjusting the penalization step of HAL regression to
respect the use of a loss function appropriate for density estimation, namely,
$L(\cdot) = -\log(g_{n,A})$~\citep{dudoit2005asymptotics}.

To build an estimator of a conditional density, \citet{diaz2011super} considered
discretizing the observed $A \in \mathcal{A}$ based on a number of bins $T$ and
a binning procedure (e.g., $T$ bins of equal length), where the choice of $T$
corresponds conceptually to the choice of bandwidth in kernel
density estimation.
To take an example, an instantiation of this procedure would divide the
observed support of $A$ into, say, $T = 7$, bins of equal length. Such
a partitioning would require $T + 1$ cutpoints along this support, yielding $T$
bins: $[\alpha_1, \alpha_2), [\alpha_2, \alpha_3), \ldots, [\alpha_6,
\alpha_7), [\alpha_7, \alpha_8]$. Next, a reformatted, repeated measures
dataset would be generated by representing each observational unit by up to $T$
records (using only $\{A_i, W_i\}_{i=1}^n$), with the number of records for
a given unit matching the rank of the bin into which a given $A_i$ falls. To
clarify, consider an individual unit $\{A_i, W_i\}$ for which the value $A_i$
falls in the fifth bin of the seven into which the support has been partitioned
(i.e., $[\alpha_5, \alpha_6)$). Five distinct records would be used to
represent the data on this single unit: $\{A_{ij}, W_{ij}\}_{j=1}^5$, where
$\{\{A_{ij} = 0\}_{j=1}^4$, $A_{i5} = 1\}$ and five exact copies of $W_i$,
$\{W_{ij}\}_{j=1}^5$.

The resultant repeated measures structure allows for the hazard probability of
$A_i$ falling in a given bin along its support to be evaluated via pooled hazard
regression (requiring only binary regression procedures). This proposal
effectively reformulates the problem into a corresponding set of hazard
regressions: $\prob (A \in [\alpha_{t-1}, \alpha_t) \mid W) = \prob (A \in
[\alpha_{t-1}, \alpha_t) \mid A \geq \alpha_{t-1}, W) \times  \prod_{j = 1}^{t
-1} \{1 - \prob (A \in [\alpha_{j-1}, \alpha_j) \mid A \geq \alpha_{j-1}, W)
\}$, where the probability of $A \in \mathcal{A}$ falling in bin
$[\alpha_{t-1}, \alpha_t)$ may be directly estimated via a binary regression
procedure, by re-expressing the corresponding likelihood in terms of the
likelihood of a binary variable in a repeated measures dataset with this
structure.
%This data augmentation procedure is carried out by creating an augmented,
%repeated measures dataset in which any given observation $A_i$ appears
%(repeatedly) for as many intervals $[\alpha_{t-1}, \alpha_t)$ as there are
%prior to the interval to which the observed $a$ belongs. A new binary outcome
%variable, indicating membership in the support set, is generated and recorded
%as part of this data restructuring process.
%With this augmented dataset, pooled hazard regression, spanning the observed
%support set of $A$, may be performed.
The hazard estimates may then be mapped to density estimates by re-scaling the
hazard estimates by the bin sizes $\lvert \alpha_t - \alpha_{t-1} \rvert$, that
is, $g_{n, \alpha}(A \mid W) = \prob(A \in [\alpha_{t-1}, \alpha_t) \mid W)
/ \lvert \alpha_t - \alpha_{t-1} \rvert$, for $\alpha_{t-1} \leq a < \alpha_t$.
We formalize this procedure in Algorithm~\ref{alg:pooled_haz_dens}.

\begin{algorithm}[h!]
\label{alg:pooled_haz_dens}
\SetAlgoLined
\KwResult{Estimates of the conditional density of $A$, given $W$.}
\SetKwInOut{Input}{Input}\SetKwInOut{Output}{Output}
\Input{
  \begin{itemize}
    \item[1] An observed data vector of the continuous treatment for $n$
      units: $A$
    \item[2] An observed data vector of the baseline covariates for
      $n$ units: $W$
    \item[3] The number of bins into which the support of $A$ is to be divided:
      $T$
    \item[4] A procedure to discretize the support of $A$ into $T$ bins:
      $\omega(\cdot)$
  \end{itemize}
}
\BlankLine

Apply $\omega(A,T)$ to divide the support of $A$ into $T$ bins:
$[\alpha_1, \alpha_2), \ldots, [\alpha_T, \alpha_{T+1}]$.

Expand the observed data in a repeated measures data structure, expressing each
unit by up to $T$ records (recording unit ID with each). For a given unit $i$,
the set of records is $\{A_{ij}, W_{ij}\}_{j=1}^{T_i}$, where $W_{ij}$ are
constant in $j$, $A_{ij}$ is a binary counting process jumping to $1$ at
$T_i$, and $T_i \leq T$ indicates the bin in which $A_i$ belongs.

Estimate, via HAL regression, the conditional hazard probability of bin
membership $\prob(A_i \in [\alpha_{t-1}, \alpha_t) \mid W)$, selecting
$\lambda_n$ by cross-validated empirical risk minimization of an appropriate
loss function~\citep{dudoit2005asymptotics}.

Rescale the conditional hazard probability estimates to the conditional density
scale by dividing the cumulative hazard by the bin widths for each $A_i$.
\BlankLine

\Output{
  $g_{n,A}$, an estimate of the generalized propensity score.
}
\caption{Pooled hazards conditional density estimation}
\end{algorithm}

Originally, a key element of this proposal was the flexibility to use any
arbitrary binary regression procedure to estimate $\prob(A \in [\alpha_{t-1},
\alpha_t) \mid W)$, facilitating the incorporation of flexible regression
techniques like ensemble modeling~\citep[e.g.,][]{breiman1996stacked,
vdl2007super}. We alter this proposal, replacing the arbitrary estimator of
$\prob(A \in [\alpha_{t-1}, \alpha_t) \mid W)$ with HAL regression, making it
possible for the resultant conditional density estimator to achieve
a convergence rate with respect to a loss-based dissimilarity of $n^{-1/4}$
under global smoothness assumptions~\citep{vdl2017generally, vdl2017uniform}, as
discussed in Section~\ref{hal}. We stress that this is a critical advance ---
essential to the asymptotic properties of our nonparametric estimators of
$\psi_{0,\delta}$.

\subsection{Nonparametric inverse probability weighted estimation}\label{npipw}

We now consider the construction of estimators of $\psi_{0,\delta}$ that rely
solely on nuisance estimation of the generalized propensity score $g_{0,A}$.
Data adaptive estimators of nuisance functionals are generally incompatible with
the direct (i.e., G-computation) and IPW estimators, as conditions necessary for
achieving asymptotic desiderata (e.g., minimal bias, efficiency) are
unattainable without imposing strong smoothness assumptions on the functional
form of the nuisance parameter estimator. This theoretical impasse has, in part,
fueled the considerable popularity enjoyed by doubly robust estimation
methodologies, such as the one-step estimation~\citep{bickel1993efficient} or
targeted minimum loss estimation frameworks~\citep{vdl2006targeted,
vdl2011targeted, vdl2018targeted}, both of which we review in the context of the
estimation of $\psi_{0,\delta}$ in Section~\ref{supp:dr_est} of the
\href{sm}{Supplementary Materials}. Briefly, doubly robust estimators require
estimation of both the outcome mechanism $\overline{Q}_{0,Y}$ and the propensity
score $g_{0,A}$; moreover, such estimators are consistent for $\psi_{0,\delta}$
when either of the two nuisance parameter estimators converge to their targets
but asymptotically efficient only when \textit{both} converge quickly enough.
When the outcome mechanism $\overline{Q}_{0,Y}$ proves challenging to estimate
well, it is tempting to consider the use of IPW estimation, which eschews
estimation of $\overline{Q}_{0,Y}$ altogether. The construction of consistent
and asymptotic efficient IPW estimators capable of incorporating data adaptive
estimation of $g_{0,A}$ has been considered before. For
example,~\citet{hirano2003efficient} proposed a logistic series estimator of
$g_{0,A}$, which requires fairly strong smoothness assumptions (i.e., $k$-times
differentiability), while~\citet{vdl2014targeted} investigated targeted
(parametric) fluctuation of $g_{n,A}$ to construct efficient IPW estimators,
which proved to be asymptotically linear but irregular. Using HAL to estimate
$g_{n,A}$,~\citet{ertefaie2020nonparametric} developed undersmoothing strategies
for asymptotically linear and efficient IPW estimation. All of these authors
exclusively considered settings with binary treatments.

We develop nonparametric-efficient IPW estimators, leveraging
Algorithm~\ref{alg:pooled_haz_dens} to construct $g_{n,A}$, and build partially
upon the contributions of~\citet{ertefaie2020nonparametric}, who demonstrated
that undersmoothing strategies could be used to select a HAL estimator of
$g_{0,A} \coloneqq \prob(A = 1 \mid W)$, ultimately yielding IPW estimators (of
the causal effects of static interventions) that achieve the nonparametric
efficiency bound.~\citet{ertefaie2020nonparametric} formulate conditions under
which $g_{n,A}$, when estimated via HAL, converges to $g_{0,A}$ at a rate
suitably fast for their nonparametric IPW estimator to achieve the asymptotic
efficiency bound in the nonparametric model. Their methods rely upon
undersmoothing the HAL-estimated $g_{n,A}$ so as to include a richer set of
basis functions than is required for optimal estimation of $g_{0,A}$, which is
achievable via $V$-fold cross-validation within the framework of loss-based
estimation~\citep{vdl2004asymptotic,dudoit2005asymptotics,vdl2006oracle}. We
propose two classes of selection procedures for undersmoothing HAL estimators of
$g_{0,A}$. These strategies share a common first step: construction of a family
of HAL-based conditional density estimators $g_{n,A}$ indexed by the
regularization term $\lambda$, i.e., $\{g_{n,A,\lambda}: \lambda_1, \ldots,
\lambda_K\}$. For this, we recommend applying
Algorithm~\ref{alg:pooled_haz_dens}, altering it to use cross-validation only to
select the strictest degree of regularization from which to begin the candidate
sequence in $\lambda$ for undersmoothing. Rather than returning a single
estimator $g_{n,A}$, the algorithm outputs $\{g_{n,A,\lambda}: \lambda_1,
\ldots, \lambda_K\}$, where $\lambda_1 \equiv \lambda_{\text{CV}}$ denotes the
degree of regularization chosen by the cross-validation selector and $\lambda_1
> \ldots > \lambda_K$ uniformly in the index set.
%We recommend the family of nonparametric estimators described by
%Algorithm~\ref{alg:pooled_haz_dens} for the very high degree of flexibility
%offered; however, the semiparametric location-scale conditional density
%estimators outlined in Algorithm~\ref{supp:alg:loc_scale_dens} can just as
%easily be adapted for this purpose, with similarly minor adjustments.
In the sequel, we assume access to a sequence of generalized propensity score
estimators $\{g_{n,A,\lambda}: \lambda_1, \ldots, \lambda_K\}$ and
a corresponding sequence of IPW estimators $\{\psi_{n,\delta,\lambda}:
\lambda_1, \ldots, \lambda_K\}$. Then, what remains is to select an IPW
estimator exhibiting desirable asymptotic properties from this sequence of
candidates.

With similar goals,~\citet{ertefaie2020nonparametric} propose two undersmoothing
criteria: (1) minimization of the mean of the EIF estimating equation up to
a theoretically desirable degree, and (2) minimization of a score term $(A
- g_{n,A})$ attributable to the treatment mechanism. Their first selector
explicitly uses the form of the EIF and must be derived anew for any given
intervention class (e.g., static vs.~stochastic). We provide the first explicit
re-characterization of the EIF~\eqref{eqn:eif_full} in a form suitable for IPW
estimator selection based on this strategy. Their second selection procedure is
incompatible with MTPs, as there is no explicit score term for the treatment
mechanism in this setting. Intuitively, this is attributable to the fact that
MTPs depend on the natural value of $A$, which is not the case for simpler
static or dynamic regimes. As an alternative, we develop a class of selection
procedures sensitive to changes along the sequence of IPW estimators
$\{\psi_{n,\delta,\lambda}: \lambda_1, \ldots, \lambda_K \}$. As part of these
proposals, we provide, to the best of our knowledge, the first demonstration of
conditional density estimator undersmoothing in causal inference.

\subsection*{Targeted undersmoothing with the influence function}

To select an IPW estimator from the sequence $\{\psi_{n, \delta, \lambda}:
\lambda_1, \ldots, \lambda_K \}$ based on the EIF, it is necessary to re-express
the EIF in a form incorporating the IPW estimating function, which can be done
by projecting this object onto the space of all functions of $A$ that are
mean-zero conditional on $W$~\citep{robins1994estimation,vdl2003unified}. The
IPW mapping defining the estimating function of the stabilized IPW
estimator~\eqref{eqn:ipw_stable} is
\begin{equation}\label{eqn:ipw_ee}
  U(O; g_A, \psi) = \frac{\tilde{g}_{A}(A \mid W)} {g_{A}(A \mid W)}
    (Y - \psi) \ .
\end{equation}
Note that the IPW estimator~\eqref{eqn:ipw_stable} is merely a solution to the
estimating function~\eqref{eqn:ipw_ee} (i.e., obtained by setting $P_n U(g_{A},
\psi) = 0$). We next present Lemma~\ref{lemma:dcar}, which characterizes the
form of the EIF estimating equation suitable for IPW estimator selection.
\begin{lemma}[IPW representation of the EIF]\label{lemma:dcar}
  Let $D^{\star}(P)$ denote the EIF~\eqref{eqn:eif_full} and let
  $U(g_{A}, \psi)$ denote the IPW estimating function~\eqref{eqn:ipw_ee}.
  Projecting $U(g_{A}, \psi)$ onto $\mathcal{F}_{\text{CAR}} =
  \{\zeta(A,W): P_0 \{\zeta(A,W) \mid W \} = 0\}$, the space of all functions
  of $A$ that are mean-zero, conditional on $W$, yields $D_{\text{CAR}}(P)$.
  %\id{Minor point, but this is not a tangent space in its strict definition.
  %A tangent space is roughly defined as the space of all scores that do not
  %change the parameter of interest. These are scores of $g$, which do change
  %the parameter of interest in the case of stochastic interventions. This is an
  %example of why I mentioned it was important to define exactly what is meant
  %by the notion coarsening at random for this kind of parameter. More
  %importantly, we do not need that notion, using IPW estimating equation minus
  %EIF would suffice. No need to change if you do not want to, as I said I will
  %not insist on this point.}
  Using $U(g_{A}, \psi)$ and $D_{\text{CAR}}(P)$, the EIF is
  $D^{\star}(P) = U(g_{A}, \psi) - D_{\text{CAR}}(P)$
  \citep{robins1994estimation,vdl2003unified}, where
  \begin{align*}
    D_{\text{CAR}}(O; P_0) = \overline{Q}_{0,Y}(A,W)
      \frac{\tilde{g}_{0,A}(A \mid W)}{g_{0,A}(A \mid W)} -
      \int_{\mathcal{A}} \overline{Q}_{0,Y}(a,W)
      \tilde{g}_{0,A}(a \mid W) d\mu(a) \ .
  \end{align*}
\end{lemma}
Among a family of IPW estimators $\{\psi_{n,\delta,\lambda}: \lambda_1,
\ldots, \lambda_K\}$, an asymptotically efficient estimator
minimizes $\lvert P_n D_{\text{CAR}}(g_{n,A,\lambda}, \overline{Q}_{n,Y})
\rvert$ for some $\overline{Q}_{n,Y}$, per Theorem~\ref{thm:ipw_eff}, which we
will formally present shortly. Such an IPW estimator is an approximate solution
to the estimated EIF as per Lemma~\ref{lemma:dcar}. The lemma's derivation is
provided in the \href{sm}{Supplementary Materials}.
%As noted in Lemma~\ref{lemma:dcar}, the term $D_{\text{CAR}}$ arises by
%projection of the estimating function $U_{g_A}(\Psi)$ onto the tangent space
%$T_{\text{CAR}} = \{\zeta(A,W): P_0 \{\zeta(A,W) \mid W \}
%= 0\}$~\citep{robins1994estimation, vdl2003unified}.
Since IPW estimators are constructed as explicit solutions to
%\id{the following expression is not an equation, and we can't talk about its
%solutions, should be $ P_n U_{g_A}(\cdot; \Psi) =0$}
$P_n U(g_{A}, \psi) = 0$, the first term of the EIF representation in the lemma
is solved; thus, selection of an asymptotically efficient IPW estimator need
only consider $D_{\text{CAR}}$.

Evaluating $D_{\text{CAR}}$ requires integration over $\mathcal{A}$, which can
prove computationally taxing on account of the need to combine techniques for
numerical integration with those for conditional density estimation. When $A$ is
discrete with known bounds, this integral expression is easily simplified;
however, when $A$ is continuous, evaluation of $D_{\text{CAR}}$ as a basis for
developing a selection criterion can be impractical. To resolve this, we
introduce the following alternative expression as a practical innovation:
\begin{equation*}
  \widetilde{D}_{\text{CAR}}(O; P) = \left[\overline{Q}^d_{Y}(A,W) -
  \left(\frac{\tilde{g}_{A}(A,W)}{g_{A}(A,W)}\right)
  \overline{Q}_{Y}(A,W)\right] -
  \Psi(P) \left(\frac{g_{A}(A,W) - \tilde{g}_{A}(A,W)}{g_{A}(A,W)}\right) \ ,
\end{equation*}
which is obtained by combining $D_{\text{CAR}}$ with another score term
appearing in the EIF; details are given in Section~\ref{supp:shift_dcar} of the
\href{sm}{Supplementary Materials}. Via this simplification,
$\widetilde{D}_{\text{CAR}}$ can be made to serve reliably as a suitable
criterion for both discrete and continuous $A$. Now, a \textit{targeted}
selector, making use of the form of $D_{\text{CAR}}$, is
\begin{equation}\label{eqn:dcar_crit}
  \lambda_n = \argmin_{\lambda} \lvert P_n
    \widetilde{D}_{\text{CAR}}(g_{n,A,\lambda},
    \overline{Q}_{n,Y}) \rvert \ .
\end{equation}
By minimizing the empirical mean of the estimated EIF absolutely, this selector
runs the risk of undersmoothing past the point at which asymptotic efficiency is
attained. Another suitable selection criterion,  proposed by
\citet{ertefaie2020nonparametric}, chooses the first $\lambda_n$ at which
$\lvert P_n \widetilde{D}_{\text{CAR}}(g_{n,A,\lambda}, \overline{Q}_{n,Y})
\rvert$ falls below $\{\sigma_{n, \lambda_{\text{CV}}} / \log(n)\}$, where
$\sigma_{n, \lambda_{\text{CV}}}$ is the standard error estimate based on the
EIF at the cross-validated choice of $\lambda$; this criterion balances
asymptotic bias against standard error while prioritizing asymptotic efficiency.

To characterize the efficiency of our proposed IPW estimators more completely,
we examine the second-order remainder $R_2$ of the EIF
expansion~\eqref{eqn:r2_expansion}. Such higher-order representations provide
insight into the behavior of efficient estimators, and an exact expression
for $R_2$ can be used to study the degree to which candidate estimators achieve
higher-order forms of efficiency. Lemma~\ref{lemma:R2} provides an expression
for this second-order term; a proof is given in Section~\ref{supp:r2_eif} of the
\href{sm}{Supplementary Materials}.
\begin{lemma}[Second-order expansion of the EIF]\label{lemma:R2}
Denote by $\{g_{n,A}, \overline{Q}_{n,Y}\}$ the nuisance estimators of
$\{g_{0,A}, \overline{Q}_{0,Y}\}$, the target nuisance functionals; then, $R_2$
is
\begin{align*}
  R_2(&g_{n,A}, \overline{Q}_{n,Y}, g_{0,A}, \overline{Q}_{0,Y}) = \\
  &\mathop{\mathlarger{\int}}\limits_{\mathcal{W}}
  \mathop{\mathlarger{\int}}\limits_{\mathcal{A}}
  \left(\overline{Q}_{n,Y}(a, w) - \overline{Q}_{0,Y}(a, w)\right)
  \left(\frac{g_{0,A}(a \mid w) - g_{n,A}(a \mid w)}{g_{n,A}(a \mid w)}
  \right) \tilde{g}_{n,A}(a \mid w) d\mu(a) d\nu(w)\ .
\end{align*}
\end{lemma}
Careful examination of Lemma~\ref{lemma:R2} reveals that the form of
$R_2$ is intimately related to that of $D_{\text{CAR}}$, which appears
in the IPW representation of the EIF in Lemma~\ref{lemma:dcar} --- in
fact, $R_2(g_{n,A}, \overline{Q}_{n,y}, g_{0,A}, \overline{Q}_{0,y}) \equiv P_0
D_{\text{CAR}}(g_{n,A}, \overline{Q}_{n,Y} - \overline{Q}_{0,Y})$. This
remainder term characterizes the contributions to asymptotic efficiency of
differences between the nuisance estimators $\{g_{n,A}, \overline{Q}_{n,Y}\}$
and the corresponding nuisance functionals $\{g_{0,A}, \overline{Q}_{0,Y}\}$. As
such, this second-order term may be used to deepen our understanding of how
improved nuisance estimation can impact estimator efficiency in a manner that
standard (first-order) EIF expansions are not equipped to capture. Notably, the
term $R_2$ cannot be evaluated in real-world settings as it requires explicit
knowledge of the forms of $\{g_{0,A}, \overline{Q}_{0,Y}\}$, which are never
available in practice; however, the form of $R_2$ itself can still be useful in
(1) deriving higher-order efficient estimators~\citep[e.g.,][]{robins2008higher}
and (2) numerically benchmarking the relative efficiency of both established and
novel estimators.

We next characterize the asymptotic properties of our undersmoothed IPW
estimators in Theorem~\ref{thm:ipw_eff}, which essentially demonstrates that IPW
estimators are asymptotically efficient under regularity and rate-convergence
conditions considered standard in semiparametric estimation. We include its
proof below since the rationale is straightforward. First, consider the
following conditions, necessary to establish Theorem~\ref{thm:ipw_eff}.

\begin{condition}[Donsker class]
  $D^{\star}(g_{n,A,\lambda}, \overline{Q}_{n,Y}) \in
  \mathcal{F}^{\star}_{v}(M)$, where $\mathcal{F}^{\star}_{v}(M)$ is a Donsker
  class and $M < \infty$ is a universal constant bounding the sectional
  variation norm.
  \label{cond:donsker_class}
\end{condition}

\begin{condition}[$D_{\text{CAR}}$ is approximately solved]
  $P_n \widetilde{D}_{\text{CAR}}(g_{n,A,\lambda_n},
  \overline{Q}_{n,Y}) = o_P(n^{-1/2})$.
  \label{cond:eif_solved}
\end{condition}

\begin{condition}[Vanishing remainder]
  $R_2(g_{n,A,\lambda_n},\overline{Q}_{n,Y}, g_{0,A}, \overline{Q}_{0,Y})
  = o_P(n^{-1/2})$.
  \label{cond:r2_vanishes}
\end{condition}

%\id{I added the below theorem along with a sketch of the proof. Feel
%free to move around either/both to whatever section of the paper you
%think is more appropriate. Feel free to also expand on the details of
%the proof if you want to, though it is not necessary.}
\begin{theorem}[Asymptotic linearity and efficiency of undersmoothed
IPW]\label{thm:ipw_eff}
Let $\lambda_n$ be an undersmoothed selection along the HAL regularization path
in $\lambda$, e.g., as in expression~\eqref{eqn:dcar_crit}, and let
$\psi_{n,\delta,\lambda_n}$ denote the solution to the estimating equation $P_n
U(g_{n,A,\lambda_n}, \psi) \approx 0$ in $\psi$. Then, under
Conditions~\ref{cond:donsker_class}--\ref{cond:r2_vanishes}, we have
\begin{equation*}
  \psi_{n,\delta,\lambda_n} - \psi_{0,\delta} =
  n^{-1} \sum_{i=1}^n D^\star(O_i;P) + o_P(n^{-1/2}) \ .
\end{equation*}
\end{theorem}
\begin{proof}
Define $\widetilde{\psi}_{n,\delta,\lambda_n}$ as the solution in $\psi$ to the
EIF estimating equation $P_n D^\star(g_{n,A,\lambda_n}, \overline{Q}_{n,Y},
\psi)$. Under Conditions~\ref{cond:donsker_class}--\ref{cond:r2_vanishes},
\begin{equation*}
  P_n D^\star(g_{n,A,\lambda_n}, \overline{Q}_{n,Y},
  \widetilde{\psi}_{n, \delta, \lambda_n}) = o_P(n^{-1/2}) \ .
\end{equation*}
Since $g_{n,A,\lambda_n}$ and $\overline{Q}_{n,Y}$ are HAL estimators,
they are both of bounded sectional variation norm with universal bound $M$,
fulfilling Condition~\ref{cond:donsker_class}~\citep{vdl2017generally}. Note
that Condition~\ref{cond:r2_vanishes} is standard in semiparametric estimation
and Condition~\ref{cond:eif_solved} is satisfied by the design of our selection
procedures. Then, applying standard arguments for doubly robust
estimation~\citep[e.g.,][]{vdl2011targeted, kennedy2016semiparametric} yields
\begin{equation*}
  \widetilde{\psi}_{n, \delta, \lambda_n} - \psi_{0, \delta} =
  n^{-1} \sum_{i=1}^nD^\star(O_i;P) + o_P(n^{-1/2}) \ .
\end{equation*}
The result follows from noticing that since both $P_n
\widetilde{D}_{\text{CAR}}(g_{n,A,\lambda_n}, \overline{Q}_{n,Y})
= o_P(n^{-1/2})$ and $D^{\star}(P) = U(g_A, \psi) - D_{\text{CAR}}(P)$, it must
also hold that $\widetilde{\psi}_{n,\delta,\lambda_n}
= \psi_{n,\delta,\lambda_n} + o_P(n^{-1/2})$.
\end{proof}
The proof establishes that an IPW estimator selected based on an undersmoothing
criterion like~\eqref{eqn:dcar_crit} will be equivalent to an asymptotically
efficient estimator under standard rate-convergence and entropy conditions on
the nuisance estimators, which HAL has been shown to satisfy. Even the entropy
conditions can be further relaxed by cross-validating the IPW
estimator~\citep{klaassen1987consistent, zheng2011cross}. While the theorem's
results require that the estimating equation $P_n
D_{\text{CAR}}(g_{n,A,\lambda_n}, \overline{Q}_{n,Y}) = o_P(n^{-1/2})$, and, by
extension, an estimating equation based on $D^{\star}$, be solved reasonably
well, we may not explicitly need to use $D_{\text{CAR}}$ as part of a selection
criterion, as we detail next.

\subsection*{Agnostic undersmoothing without the influence function}

Explicit (re)characterization of the EIF for the selection of an IPW estimator
can prove a challenging endeavor, requiring specialized and tedious mathematical
manipulations. As such, it is often preferable to select an IPW estimator in
a manner agnostic to the EIF's form. Firstly, such agnostic selection procedures
are better aligned philosophically with IPW estimation, which avoids explicit
modeling of the outcome mechanism (n.b., the EIF always contains a term for
$\overline{Q}_{0,Y}$). Secondly, agnostic procedures may solve score equations
beyond those explicitly appearing in the EIF expression, allowing them to
automatically and blindly satisfy higher-order efficiency desiderata.
%A procedure that does not make use of the EIF has the additional advantage of
%remaining applicable across a wide range of intervention regimes, allowing for
%its use in a possibly vast array of settings, without the need for either
%re-derivation or re-implementation.
%\id{I know where you are coming from with this, but a reader may not understand
%that you have in mind a general application of this methodology to estimation
%of any parameter, because you have not discussed that yet.}
We formulate two selection procedures that do not use the form of the EIF,
instead considering properties of the IPW estimators $\{\psi_{n,\delta,
\lambda}: \lambda_1, \ldots, \lambda_K \}$ along their regularization trajectory
$\{\lambda_1, \ldots, \lambda_K\}$. In the context of nonparametric sieve
estimation, such ideas were formulated in the seminal work
of~\citet{lepskii1992asymptotically, lepskii1993asymptotically}, with sporadic
extensions since~\citep[e.g.,][]{davies2014sieve, mukherjee2015lepski,
vdl2018cvtmle, cai2019nonparametric, qiu2021universal}.

The aim of formulating such EIF-\textit{agnostic} undersmoothing selection
procedures is to produce selections similar to (or better than) those of the
targeted selectors. For IPW estimators based on these agnostic selectors to
asymptotically attain the nonparametric efficiency bound, these estimators must
solve the EIF estimating equation. Our first proposal in this class balances
changes along the regularization sequence $\{\lambda_1, \ldots, \lambda_K \}$ in
the IPW estimator $\psi_{n,\delta,\lambda}$ against changes in its estimated
standard error $\sigma_{n,\lambda}$. This is merely the first element in the
sequence $\{\lambda_1, \ldots, \lambda_K\}$ for which the condition
\begin{equation}\label{eqn:plateau_var_cond}
 \lvert \psi_{n,\delta,\lambda_{j+1}} - \psi_{n,\delta,\lambda_j}
 \rvert \leq \log(n)^{-1} Z_{(1-\alpha/2)}
 \lvert \sigma_{n,\lambda_{j+1}} - \sigma_{n,\lambda_j}\rvert \quad
 \text{for}~j = \{1, \ldots, K-1\}
\end{equation}
is met. Note that $Z_{(1-\alpha/2)}$ is the $(1-\alpha/2)$\textsuperscript{th}
quantile of the standard Gaussian distribution. While we recommend the use of
the stabilized IPW estimator~\eqref{eqn:ipw_stable} for
$\psi_{n,\delta,\lambda}$, there are several candidate choices for the standard
error estimator $\sigma_{n,\lambda}$. We recommend basing this on the
empirical variance of the \textit{estimated} IPW estimating function $n^{-1}
\{\mathbb{V}[U(g_{n,A}, \psi_{n,\delta,\lambda})]\}$.
While~\citet{haneuse2013estimation} demonstrated this variance estimator to be
anti-conservative (contrary to naive theoretical expectations), this choice is
made for the sake of computational ease, as only the successive differences in
standard error estimates along the regularization path are necessary.
%--- that is, we do \textit{not} recommend using this estimator for
%inference on $\psi_{n,\delta,\lambda}$ at the selected $\lambda_n$.
An alternative variance estimate could be obtained via the nonparametric
bootstrap, the functional smoothness conditions of which HAL has been shown to
satisfy~\citep{cai2019nonparametric}. Of course, $\sigma_{n,\lambda}$ could also
be based on the empirical variance of the EIF, though this obviously undermines
the selector's design goal of EIF-agnosticism. We stress that our choice of
variance estimator is imperfect (possibly failing to reflect increases in
variance that accompany the relaxation of regularization), but it provides
a reasonable surrogate for variance changes that balances computational
expedience and inferential stability. Examination of
equation~\eqref{eqn:plateau_var_cond} reveals that this selector identifies
$\lambda_n$ as the first point in the sequence $\{\lambda_1, \ldots,
\lambda_K\}$ at which the relaxation of regularization impacts the IPW point
estimates less than it affects their corresponding standard error estimates,
indicating that a desirable bias-variance tradeoff has been achieved. The
scaling of the standard error estimate by $Z_{(1-\alpha/2)} / \log(n)$ ensures
that changes in the point estimate are weighed against changes in confidence
interval width, a surrogate for inferential efficiency.
%and that MSE is optimized
%asymptotically. Since this agnostic selection procedure draws inspiration from
%the developments of~\citet{lepskii1991problem, lepskii1992asymptotically,
%lepskii1993asymptotically}, as homage, we refer to it as ``Lepski's plateau
%selector'' in the sequel.

A potential pitfall of the immediately preceding proposal is its requirement of
standard error estimation, which can be computationally challenging (e.g.,
requiring the bootstrap), result in unstable or unreliable estimates (e.g., when
the standard error estimator is itself poor), or require nuisance estimation
beyond $g_{0,A}$ (i.e., when the standard error estimated is based on the EIF).
Thus, such an agnostic selector may, at times, fail to choose the optimal IPW
estimator among a sequence, especially in scenarios in which standard error
estimation proves challenging. It is possible to eschew such requirements
altogether, relying instead entirely on the trajectory of
$\{\psi_{n,\delta,\lambda}: \lambda_1, \ldots, \lambda_K \}$ alone. Under such
a strategy, the selection $\lambda_n$ would simply be the first term in the
sequence $\{\lambda_1, \ldots, \lambda_K\}$ at which the trajectory taken by
$\{\psi_{n,\delta,\lambda}: \lambda_1, \ldots, \lambda_K \}$ encounters
a plateau. We formalize this second class of agnostic selection procedures in
Algorithm~\ref{alg:plateau_psi}.

\begin{algorithm}[h!]
\label{alg:plateau_psi}
\SetAlgoLined
\KwResult{A nonparametric-efficient IPW estimator $\psi_{n,\delta,\lambda_n}$
of $\psi_{0,\delta}$.}
\SetKwInOut{Input}{Input}\SetKwInOut{Output}{Output}
\Input{
  \begin{itemize}
    \item[1] A regularization sequence in decreasing order:
      $\{\lambda_1, \ldots, \lambda_K\}$
    \item[2] A corresponding sequence of IPW estimators:
      $\{\psi_{n,\delta,\lambda}: \lambda_1, \ldots, \lambda_K\}$
    \item[3] A corresponding sequence of HAL estimators of $g_{0,A}$:
      $\{g_{n,A,\lambda}: \lambda_1, \ldots, \lambda_K\}$
    \item[4] A positive scalar of the maximum multiplier of the CV-chosen
      $L_1$-norm: $K_{\text{max}}$
    \item[5] An algorithm to detect inflection points in a sequence:
      $\tau(\cdot)$

  \end{itemize}
}
\BlankLine

Estimate $L_1$-norm for each IPW estimator in the given sequence by summing the
coefficients for the nuisance estimators $\{g_{n,A,\lambda}: \lambda_1, \ldots,
\lambda_K\}$, denoting this $\{M_{n,\lambda_1}, \ldots, M_{n,\lambda_K}\}$. The
$L_1$-norm of the CV-selected estimator $g_{n,A,\lambda_1}$ is
$M_{n,\lambda_1}$.

Select a window of candidate IPW estimators anchored in the neighborhood of the
CV-selected estimator $\psi_{n,\delta,\lambda_1}$ by computing a confidence
interval around it as $\{\psi_{n,\delta, \lambda_1} \pm Z_{1-\alpha/2}
\sigma_{n,\lambda_1}\}$, with $\sigma_{n,\lambda_1}$ being the standard error
estimate based on $U(g_{n,A,\lambda_n}, \psi)$ and $Z_{1-\alpha/2}$ being
a standard Gaussian quantile, and, then, setting the $L_1$-norm's maximum
allowed value as $M_{n,\text{max}} \coloneqq K_{\text{max}} M_{n,\lambda_1}$.
Reduce the neighborhood of candidate estimators to $\{\psi_{n,\delta,\lambda}:
\lambda_1, \ldots, \lambda_J\}$ for $J < K$ falling within the confidence region
and with $L_1$-norms no greater than $M_{n,\text{max}}$.

Construct a smoothed trajectory of the IPW estimators
$\{\psi_{n,\delta,\lambda}: \lambda_1, \ldots, \lambda_J\}$ against the
subsequence of $L_1$-norms by applying LOESS (or similar).

Find an inflection point along the smoothed trajectory by applying
$\tau(\cdot)$. If this succeeds, denote the estimator at this index
$\psi_{n,\delta}$. If $\tau(\cdot)$ fails, let $\psi_{n,\delta} \coloneqq
\psi_{n,\delta,\lambda_1}$.

\BlankLine

\Output{
  $\psi_{n,\delta}$, a nonparametric IPW estimate of $\psi_{0,\delta}$.
}
\caption{Smoothed plateau selector for nonparametric IPW estimation}
\end{algorithm}

Such a selection procedure considers the ``sharpness'' of changes in the point
estimates $\psi_{n,\delta,\lambda}$ sequentially in $\{\lambda_1, \ldots,
\lambda_K\}$ as the degree of regularization is relaxed. This selector aims to
identify a value $\lambda_n$ at which the trajectory of the IPW point estimate
$\psi_{n,\delta,\lambda_n}$ encounters an inflection point, a phenomenon
suggesting that further relaxations of the regularization parameter ought not
contribute meaningfully to the goal of debiasing the IPW estimator through
undersmoothing. As this trajectory may be insufficiently smooth (e.g., when
$\{\lambda_1, \ldots, \lambda_K\}$ is too coarse) to reveal such inflection
points, LOESS can be leveraged to enforce a sufficient degree of smoothness,
helping a given inflection-finding algorithm escape failure. This agnostic
selection algorithm attempts to find an inflection point in the trajectory of
the IPW estimators only within a limited neighborhood beyond the IPW estimator
chosen by cross-validation $g_{n,A,\lambda_{\text{CV}}}$. This form of anchoring
serves to narrow the set of allowable IPW estimators considered by the selector
such that the degree of debiasing enforced by undersmoothing is not overly
extreme, as undersmoothing uncontrollably could lead to the inclusion of HAL
basis functions irrelevant for optimal estimation of $\psi_{0,\delta}$.
Intuitively, this form of anchoring ensures that this selection procedure
operates in a neighborhood of $L_1$-norm values roughly similar to those
considered by the targeted selection procedures previously formulated, a claim
inspired and corroborated by extensive numerical experiments conducted during
this algorithm's development. In those rare instances in which an inflection
point cannot be found within this neighborhood, the selection algorithm simply
returns the IPW estimator based on $g_{n,A,\lambda_{\text{CV}}}$.
%Drawing inspiration from UCB-style algorithms popular in the literature on
%multi-armed bandits, in those rare instances in which an inflection point
%cannot be found within this neighborhood, the selection algorithm returns
%a point estimate as the upper confidence bound (i.e., UCB) of the IPW estimator
%based on $g_{n,A,\lambda_{\text{CV}}}$.
We conjecture that an IPW estimator identified by this selection algorithm will
suitably solve the EIF estimating equation and thus achieve desirable asymptotic
properties, though we leave the theoretical formalization of this claim as
a topic of future investigation.
%\nh{thoughts, Mark?}

%Lemma~\ref{lemma:plateau} characterizes this behavior.

%\ah{Any heuristics as to why this might work? Seems sensible for sure.}

%Lemma 2: Plateau selection satisfies the EIF [TO DO]
%\begin{lemma}[Plateau selection satisfies the EIF]\label{lemma:plateau}
  %To-do
%\end{lemma}

%%%%%%%%%%%%%%%%%%%%%%%%%%%%%%%%%%%%%%%%%%%%%%%%%%%%%%%%%%%%%%%%%%%%%%%%%%%%%%%
\section{Numerical studies}\label{sim}

We evaluated our proposed IPW estimators in simulation experiments using two
distinct data-generating processes (DGPs) for the treatment mechanism. These
experiments aimed to tease apart the impact treatment mechanism's form on the
performance of our IPW estimators relative to an alternative class of doubly
robust (DR) estimators popular in modern causal inference. Specifically, we
compared our IPW estimators to targeted minimum loss estimators of
$\psi_{0,\delta}$ (reviewed briefly in Section~\ref{supp:dr_est} of the
\href{sm}{Supplementary Materials}), as these estimators can incorporate
flexible regression strategies for nuisance estimation. We exclude from our
comparison classical IPW estimators based on parametric estimation of $g_{0,A}$,
as such alternatives are prone to model misspecification bias and generally fail
to achieve the nonparametric efficiency
bound~\citep{vdl2003unified,ertefaie2020nonparametric}. Across all reported
experiments, we consider IPW estimators over a regularization trajectory of
$3000$ preset candidate values of $\lambda$. We demonstrate that undersmoothing
of $g_{n,A}$ (based on Algorithm~\ref{alg:pooled_haz_dens}) succeeds in
debiasing our IPW estimators of $\psi_{0,\delta}$, conferring benefits in terms
of both bias and asymptotic efficiency.
%Both the estimator of Algorithm~\ref{alg:pooled_haz_dens} and our novel IPW
%estimators are implemented in the \texttt{haldensify} \texttt{R}
%package~\citep{hejazi2021haldensify}.

Throughout our experiments, we compare several variants of our IPW estimators
$\psi_{n,\delta,\lambda_n}$, differing by the manner in which $\lambda_n$ is
selected. We consider a total of six estimator variants, including (1) one in
which $\lambda_n$ is selected by ``global'' cross-validation (CV), which is
optimal for estimation of $g_{0,A}$ but not $\psi_{0,\delta}$; (2) two based on
the targeted criteria of Lemma~\ref{lemma:dcar}, including~\eqref{eqn:dcar_crit}
and its tolerance-based alternative~\citet{ertefaie2020nonparametric}; (3) one
based on the agnostic selection criterion~\eqref{eqn:plateau_var_cond}; (4)
another based on the agnostic selection approach of
Algorithm~\ref{alg:plateau_psi}; and (5) a hybrid selector combining the
agnostic selection approach of Algorithm~\ref{alg:plateau_psi} with a targeted
stopping point, which limits its candidates to regions along the regularization
grid prior to the $\lambda_n$ selected by criterion~\eqref{eqn:dcar_crit}. From
among these six selectors, two --- the ``global'' CV selector and the targeted
selector~\eqref{eqn:dcar_crit} minimizing $\lvert P_n
\widetilde{D}_{\text{CAR}}(g_{n,A,\lambda}, \overline{Q}_{n,Y}) \rvert$ --- are
used as candidates for $g_{n,A}$ in the DR estimators, which use either HAL
regression or a (misspecified) GLM to estimate $\overline{Q}_{n,Y}$. Here, we
compare our estimators in terms of scaled bias (multiplied by $n^{1/2}$), scaled
mean-squared error relative to the efficiency bound of the given DGP, and the
degree to which an estimator solves the empirical mean of the EIF. In
Section~\ref{supp:addl_sim_res} of the \href{sm}{Supplementary Materials}, we
examine estimator performance in terms of (1) the coverage of oracle confidence
intervals, which use a Monte Carlo variance estimator (eschewing any problems in
variance estimation); (2) the coverage of Wald-style confidence intervals, which
use the empirical variance of the EIF (and may be prone to poor variance
estimation); and (3) an approximated second-order remainder $R_2$ of the EIF.

For each DGP, we consider a collection of observed data units $O = (W_1, W_2,
W_3, A, Y)$, where $W_1 \sim \text{Bern}(p = 0.6)$, $W_2 \sim \text{Unif}(\min
= 0.5, \max = 1.5)$, and $W_3 \sim \text{Pois}(\lambda = 2)$; the generating
functions for $A$ and $Y$ vary across the two scenarios. For each simulation
experiment, in a given scenario, we sample $n \in \{100, 200, 500\}$ units and
apply the IPW and DR estimator variants to estimate $\psi_{0,\delta}$ at
$\delta = 1$, repeating each experiment $300$ times. We approximate
$\psi_{0,\delta}$ in each scenario by applying the direct
estimator~\eqref{eqn:plugin}, using the known $\overline{Q}_{0,Y}$, in a single,
large sample of $n = 1,000,000$. For the evaluation criteria considered, the
scaled asymptotic bias (i.e., $n^{1/2}(\psi_{0,\delta}
- \psi_{n,\delta,\lambda_n})$) is expected to decrease with increasing sample
size for consistent estimators, while the scaled mean-squared error (i.e., $n
\{(\psi_{0,\delta} - \psi_{n,\delta,\lambda_n})^2 + \sigma_{n,\lambda_n}^2\}$),
relative to the efficiency bound, should converge to unity for regular efficient
estimators. All experiments were performed with version 4.0.3 of the \texttt{R}
language and environment for statistical computing~\citep{R}.

\subsection{Simulation \#1: Treatment mechanism with equal mean and
  variance}\label{hese_sim_poisson}

\begin{figure}[h!]
  \centering
  \includegraphics[scale=0.28]{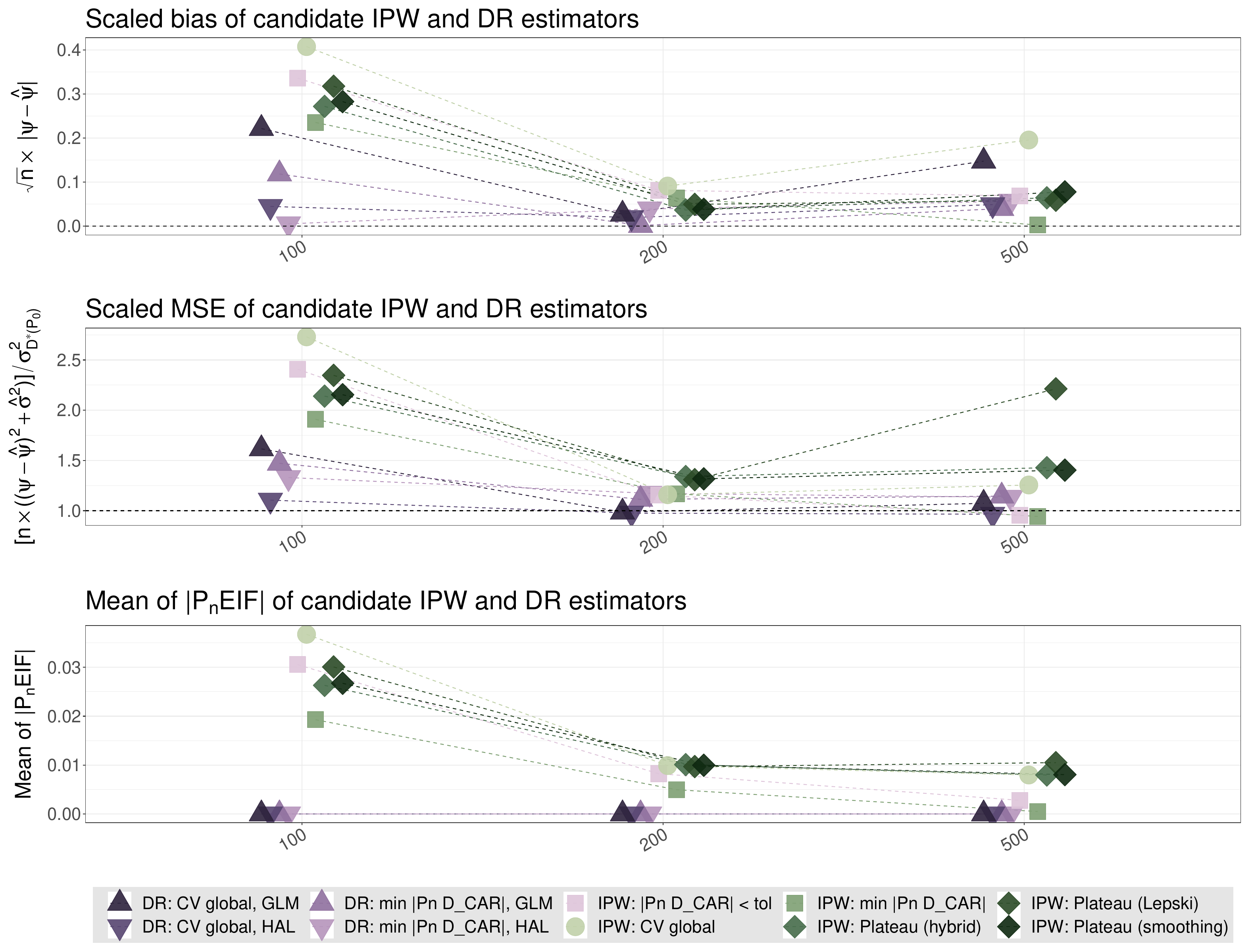}
  \caption{Numerical comparisons of our nonparametric IPW estimator variants
    and common DR estimators.}
  \label{fig:dgp2a_npipw}
\end{figure}

This setting is formulated based on a treatment mechanism governed by a Poisson
distribution, in which the mean and variance of $A$ are both equally impacted by
a subset of the baseline covariates $\{W_1, W_2\}$; the exact form of the
treatment mechanism results in $A$ taking on discrete values in a large
range. This design choice ensures that $A$ is incompatible with alternative
(static or dynamic) intervention schemes but compatible with MTPs.
%and allows it to remain compatible with the generalized propensity score
%estimator of Algorithm~\ref{alg:pooled_haz_dens}, which relies upon
%discretization of $A$ for constructing the estimator $g_{n,A}$.
%Importantly, note that incompatibility with Algorithm~\ref{alg:pooled_haz_dens}
%would lead to severe complications in attempting to distinguish bias in the
%estimation of $g_{0,A}$ from downstream bias in the estimation of
%$\psi_{0,\delta}$.
For this scenario, the treatment mechanism takes the form $A \mid W \sim
\text{Poisson}\left(\lambda = (1 - W_1) + 0.25 W_2^3 + 2 W_1 W_2 + 4\right)$ and
the outcome mechanism is $Y \mid A, W \sim \text{Bernoulli}\left(p
= \text{expit}(A + 2 (1 - W_1) + 0.5 W_2 + 0.5 W_3 + 2 W_1 W_2 - 7)\right)$.
Numerical evaluation of the performance of our proposed IPW estimators of
$\psi_{0,\delta}$ across the $300$ simulation experiments is summarized in
Figure~\ref{fig:dgp2a_npipw}.

In terms of scaled bias, all of the undersmoothed IPW estimator variants (square
and diamond markers) are comparable at $n = 200$ and $n = 500$, with performance
indistinguishable from that of the better-performing DR estimators (triangle
markers). At $n = 100$, the IPW estimators exhibit a limited degree more bias
than their DR counterparts, though this is on the order of thousandths on the
raw bias scale. The IPW estimator based on the ``global'' CV selector (circle
marker) exhibits the highest degree of bias at $n = 100$ and $n = 500$, and its
scaled bias does not appear to be converging to zero with increasing sample
size. In terms of asymptotic efficiency, the two targeted undersmoothed IPW
estimator variants (square markers) outperform the others at $n = 500$. Among
the three agnostic undersmoothed IPW estimators (diamond markers), the two based
on variations of Algorithm~\ref{alg:plateau_psi} outperform the IPW estimator
based on Lepski's selection approach~\eqref{eqn:plateau_var_cond}; however, none
of these IPW estimators achieve the efficiency bound by $n = 500$. Notably, the
IPW estimator based on criterion~\eqref{eqn:plateau_var_cond} displays inflated
variance at $n = 500$, which we conjecture to be due to occasionally poor
selections driven by unstable variance estimation, as noted previously.
Interestingly, and unexpected by theory, the IPW estimator based on the
``global'' CV selector nearly achieves the efficiency bound at $n = 200$ and $n
= 500$ but displays nontrivial bias at the larger sample size, suggesting that
its efficiency arises from reduced variance, rather than the debiasing induced
by undersmoothing. All of the estimator candidates appear to solve the empirical
mean of the EIF quite well, suggesting that all of these estimators are
asymptotically efficient by way of their being solutions to the EIF estimating
equation (up to a reasonable degree). Overall, the results of this set of
numerical experiments speak in favor of our proposed estimators, suggesting that
most, if not all, of our nonparametric IPW estimators are consistent and
efficient at even moderate sample sizes. Notably, the DR estimators achieve
better performance at $n = 100$ and should be preferred when the number of study
units available is so limited.

\subsection{Simulation \#2: Treatment mechanism with unequal mean and
  variance}\label{hese_sim_negbinom}

\begin{figure}[h!]
  \centering
  \includegraphics[scale=0.28]{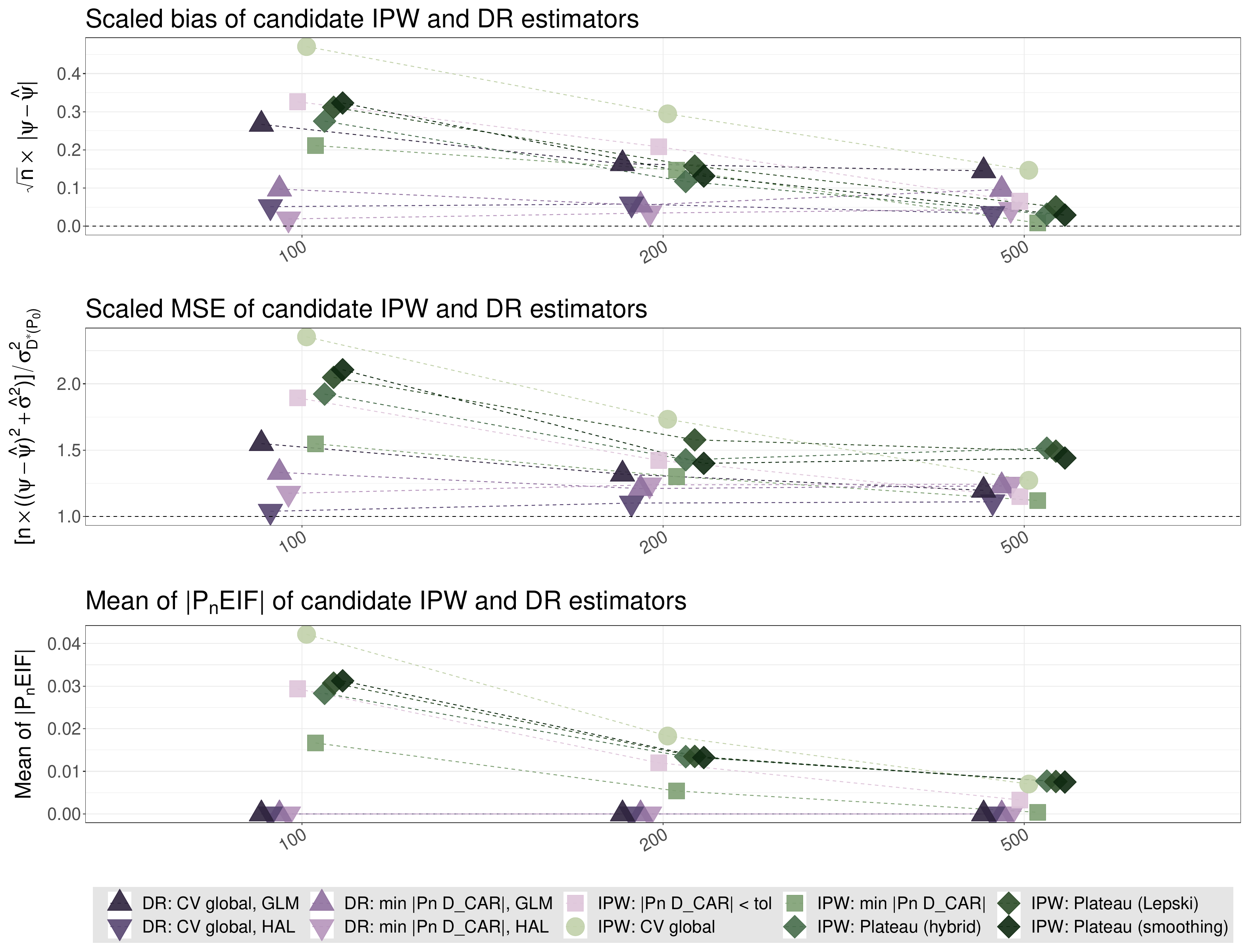}
  \caption{Numerical comparisons of our nonparametric IPW estimator variants
    and common DR estimators.}
  \label{fig:dgp2b_npipw}
\end{figure}

This experimental scenario was constructed using a treatment mechanism governed
by a Negative Binomial distribution, in which the mean $\mu$ and dispersion
$\nu$ are both impacted by a subset of the baseline covariates $\{W_1, W_2\}$.
Recall that the variance of a Negative Binomial random variable is $\mu + \mu^2
/ \nu$, so that, unlike in the previous scenario, the mean and variance of $A$
need not match.
%As the form of the treatment mechanism results in $A$ taking on a large range
%of discrete values, we expect the discretization procedure built into
%Algorithm~\ref{alg:pooled_haz_dens} to cause no significant approximation
%issues, so that the procedure's estimation quality can be disentangled from any
%bias introduced by discretization.
In this scenario, the treatment mechanism is $A \mid W \sim \text{NB}\left(\mu
= (1 - W_1) + 0.25 W_2^3 + 2 W_1 W_2 + 4, \nu = 5 W_2 + 7\right)$ while the
outcome mechanism is $Y \mid A, W \sim \text{Bernoulli}\left(p = \text{expit}(A
+ 2 (1 - W_1) + W_2 - W_3 + 1.5 W_1 W_2 - 5)\right)$. Numerical evaluation of
the performance of our proposed IPW estimators of $\psi_{0,\delta}$ across the
$300$ simulation experiments is summarized in Figure~\ref{fig:dgp2b_npipw}.

In terms of scaled bias, the undersmoothed IPW estimators (square and diamond
markers) outperform the IPW estimator based on ``global'' CV at $n = 100$ and
$n = 200$, depicting the debiasing enforced by undersmoothing. These novel IPW
estimators display performance comparable to that of the DR estimators across
all of the sample sizes considered (n.b., that the unscaled bias is on the order
of thousandths) and are indistinguishable from one another by $n = 500$. None of
the estimator variants considered appear to show any sign of being
asymptotically inconsistent. As regards asymptotic efficiency, the DR estimator
variants outperform their IPW competitors at $n = 100$ but show comparable
efficiency by $n = 200$; moreover, all estimators appear to be approaching the
efficiency bound by $n = 500$, though the IPW estimators based on the targeted
undersmoothing seem to be converging more quickly (in $n$) than their
counterparts based on agnostic undersmoothing criteria. Unlike in the preceding
scenario, the IPW estimator using agnostic undersmoothing based on
criterion~\eqref{eqn:plateau_var_cond} does not show an uptick in efficiency at
$n = 500$ but instead appears to be converging to the efficiency bound. The IPW
estimator based on ``global'' CV exhibits inflated bias and poorer efficiency
than its counterparts at $n = 100$ and $n = 200$, though its performance vastly
improves (to the point of being comparable) by $n = 500$; we stress that this is
not expected from theory and should not be taken as a general property of IPW
estimators based on such a selector. All of the IPW and DR estimator variants
appear to solve the empirical mean of the EIF quite well, suggesting that all
are solutions to the EIF estimating equation and, as such, capable of achieving
asymptotic efficiency. Of note, with respect to this metric, the IPW estimators
based on targeted undersmoothing criteria (square markers) exhibit performance
closer to that of the DR estimators than do their IPW counterparts based on
agnostic undersmoothing (diamond markers). As with the preceding scenario, the
results of this set of numerical experiments suggest that all of our novel IPW
estimators are asymptotically consistent and efficient, with performance
comparable to that of DR estimators in sample sizes as low as $n = 500$. While
the conclusion may remain that DR estimators ought to be preferred when sample
sizes are very limited (e.g., $n = 100$), our novel IPW estimators are
competitive, as measured by standard desiderata from semiparametric efficiency
theory, with access to only modest sample sizes. Furthermore, as evidenced by
the performance of the IPW estimators based on agnostic undersmoothing, our
proposed estimators can achieve such desirable levels of performance without any
explicit knowledge of semiparametric efficiency theory.

\section{Discussion}\label{discuss}

We have developed nonparametric-efficient IPW estimators of the causal effects
of MTPs by applying targeted or agnostic sieve estimation techniques to flexible
estimators of the generalized propensity score. Our IPW estimators rely entirely
on the accuracy and rate-convergence properties of the generalized propensity
score estimator; moreover, our targeted selectors rely upon estimation of the
outcome mechanism as well, which plays a role in the quality of the selection.
There are several avenues along which our proposed estimators and selection
procedures may be evaluated. For instance, further investigations may focus on
evaluating the degree to which selection of an optimal undersmoothed estimator
$g_{n,A,\lambda_n}$ is impacted by possibly poor estimation of
$\overline{Q}_{n,Y}$ as well as the degree to which the efficiency of our IPW
estimators degrades under poor estimates of the latter quantity. An analytic
strategy in this vein could lead to a sensitivity analysis for this class of IPW
estimators. Examining how our successful application of sieve estimation
principles in this setting could be expanded to other areas of causal inference
reveals further avenues for potential progress, both theoretically and
practically.

Although the formulation and demonstration of our proposed methodology in the
context of single timepoint interventions appears promising, extensions to
longitudinal intervention regimes require significant and careful attention.
\citet{diaz2020nonparametric} have already extended the theoretical underpinning
of MTPs to the longitudinal setting, though their estimation efforts focused
exclusively on sequentially doubly robust estimators. As these authors have
already derived the EIF with respect to the nonparametric model $\M$, our
efforts could focus instead on deriving the relevant $D_{\text{CAR}}$ projection
term necessary for using undersmoothing techniques to construct
nonparametric-efficient IPW estimators of the causal effects of longitudinal
MTPs. The successful formulation of sieve-based estimation strategies in this
setting promises unique challenges, as the generalized propensity score
estimators at each timepoint would need to be \textit{jointly} undersmoothed
with the aim of solving the EIF adequately \textit{across} all timepoints. No
similar strategy appears to have been previously explored in the context of IPW
estimation.

Our examinations aimed only to develop nonparametric-efficient IPW estimators of
the causal effects of MTPs, yet flexible and accurate generalized propensity
score estimators are critical for formulating doubly robust estimators as well.
In the \href{sm}{Supplementary Materials}, we presented evidence from numerical
studies suggesting that our agnostic selectors, and one of our targeted
selectors, lead to estimators that minimize the second-order remainder
associated with the EIF just as well as conventional doubly robust techniques
do. An important area of future work concerns how these selectors can be
combined with doubly robust estimators so as to develop procedures that are
doubly robust (in terms of consistency) but that yet achieve improved efficiency
and rate-convergence properties via undersmoothing of the generalized propensity
score. Whether such estimators could be made to exhibit higher-order forms of
asymptotic efficiency --- and the degree to which our selection procedures would
need to be adapted to this end --- remains an open question, though
investigations outside of the framework of sieve estimation have shown
promise~\citep[e.g.,][]{mukherjee2017semiparametric, vdl2021higher}.

%%%%%%%%%%%%%%%%%%%%%%%%%%%%%%%%%%%%%%%%%%%%%%%%%%%%%%%%%%%%%%%%%%%%%%%%%%%%%%%
%\section*{Supplementary material}

%Supplementary material available online includes proofs of presented results.

\subsection*{Acknowledgements}

We thank Alan Hubbard and Nick Jewell, for helpful comments on an early draft
of the manuscript; Jeremy Coyle, for discussions and early-stage collaborations
on software implementations; and Ted Westling, for bringing to our attention
semiparametric conditional density estimation techniques for location-scale
families. NSH's work was supported in part by the National Science Foundation
(award no.~DMS 2102840), and MJvdL was partially supported by a grant from the
National Institute of Allergy and Infectious Diseases (award no.~R01 AI074345).

%%%%%%%%%%%%%%%%%%%%%%%%%%%%%%%%%%%%%%%%%%%%%%%%%%%%%%%%%%%%%%%%%%%%%%%%%%%%%%%
\bibliography{refs}
\end{document}

\maketitle

%%%%%%%%%%%%%%%%%%%%%%%%%%%%%%%%%%%%%%%%%%%%%%%%%%%%%%%%%%%%%%%%%%%%%%%%%%%%%%%

We provide details on our theoretical arguments in Sections~\ref{shift_dcar}
and~\ref{r2_eif}, throughout which we rely on notation and quantities introduced
within the entirety of the \href{paper}{main manuscript}. In these sections, we
will, at times, make use of a convenient identity for swapping $\tilde{g}_{A}(A
\mid W)$ with $\overline{Q}_{Y}^{d}(A, W)$, itself based on the
change-of-variable formula for integration. For an arbitrary function $f(A,W)$
of $(A,W)$, this identity is
\begin{equation}\label{eqn:mtp_identity}
  \int_{\mathcal{A}} f(a, W) \tilde{g}_{A}(a \mid W) d\mu(a) =
  \int_{\mathcal{A}} f(d(a,W;\delta), W) g_A(a \mid W) b'(a,W;\delta) d\mu(a),
\end{equation}
where $d(A,W)$ is an MTP following Assumption~\ref{main:ass:piece_inv}, as in
equations~\eqref{main:eqn:shift_additive} and~\eqref{main:eqn:shift_mult}, and,
in an abuse of notation, the term $b'(A,W;\delta)$ is the derivative of the
inverse of $d(A,W;\delta)$, meant to refer to the appropriate inverse of its
primitive, for $d(A,W;\delta)$ defined piecewise. Examination of
equation~\eqref{eqn:mtp_identity} reveals that the right-hand side of the
equivalence is far easier to evaluate, since it requires only integration with
respect to $g_A(A \mid W)$, the natural conditional treatment density, instead
of integration with respect to $\tilde{g}_{A}(A \mid W)$, the counterfactual
conditional treatment density. When the MTP is as
in~\eqref{main:eqn:shift_additive}, the expression on the right-hand side
further simplifies, as $b'(A,W;\delta) = 1$ uniformly in the support of
$d(A,W;\delta)$; however, the expression is not much more complicated when the
MTP is as in~\eqref{main:eqn:shift_mult}, in which case $b'(A,W;\delta)
= \delta^{-1}$. Throughout the remainder of the present treatment, for
simplicity of exposition, we will assume that $d(A,W;\delta)$ is the
MTP~\eqref{main:eqn:shift_additive}, though similar arguments hold for
alternative MTP definitions.

\section{Developing a Targeted Undersmoothing Criterion}\label{shift_dcar}

We now turn to deriving the projection component $D_{\text{CAR}}(P_0)$ of the
EIF of $\psi_{0,\delta}$, leading to Lemma~\ref{main:lemma:dcar} presented in
Section~\ref{main:npipw} of the manuscript. We use this form of the projection
$D_{\text{CAR}}(P_0)$ to detail the construction of our pragmatic criterion
$\widetilde{D}_{\text{CAR}}(P_0)$.

Recall that we consider MTPs of the forms~\eqref{main:eqn:shift_additive}
and~\eqref{main:eqn:shift_mult}, which imply differing post-intervention
conditional treatment densities, both of which we denote as $\tilde{g}_{0, A}(A
\mid W)$. These correspond to stochastic interventions of the second variety
discussed in the main text, in which the MTP $d(A, W; \delta)$ shifts the
natural value of the treatment $A$ by an investigator-supplied scalar
$\delta(W)$, which may depend on baseline covariates $W$. To begin, we remind
ourselves that $D^{\star}(P) = U(g_A, \psi) - D_{\text{CAR}}(P)$ at
a distribution $P \in \mathcal{M}$ by a representation attributed
to~\citet{robins1994estimation}. Specifically,
%criterion the EIF representation based on the projection of the IPW estimating
%function $D_{\text{IPW}}$ onto the tangent space of coarsened-at-random
%functions $D_{\text{CAR}}$. Let's note that the EIF $D^{\star}$ may be expressed
%as follows:
\begin{align*}
  D^{\star}(O; P) =&
    \underbrace{\left(Y \frac{\tilde{g}_{A}(A \mid W)}
      {g_{A}(A \mid W)} - \Psi(P) \right)}_{U(g_A, \psi)} \\
    &- \underbrace{\left(\overline{Q}_{Y}(A,W)
      \frac{\tilde{g}_{A}(A \mid W)}{g_{A}(A \mid W)} -
    \int_{\mathcal{A}} \overline{Q}_{Y}(a,W)
      \tilde{g}_{A}(a \mid W) d\mu(a) \right)}_{\text{score for a
      \textit{known} stochastic intervention ($D_{\text{CAR}}$)}} \\
    &+ \underbrace{\left(\overline{Q}_{Y}^{d}(A, W) -
      \E[\overline{Q}_{Y}^{d}(A, W) \mid W]
      \right)}_{\text{score for a \textit{learned} stochastic
      intervention ($s_g$)}} \ .
\end{align*}
Note that this general representation of the EIF accounts for the dependence of
the post-intervention treatment density $\tilde{g}_{A}(A \mid W)$ on the
distribution $P$, that is, where the target parameter is the counterfactual mean
of $Y$ under a stochastic intervention of the first variety discussed in the
main text. While this general representation encompasses stochastic
interventions that do not strictly satisfy Assumption~\ref{main:ass:piece_inv},
required of MTPs, it is challenging to work with on account of both the second
term, which requires integration over $\mathcal{A}$, and the third term, which
involves a score based on the outcome mechanism $\overline{Q}_{Y}$. Fortunately,
when the stochastic intervention is an MTP (as in the scope of this manuscript),
a significant simplification can be made. Namely, when the post-intervention
conditional density is ``known'' (i.e., implied by an MTP $d(A, W; \delta)$),
the dependence of the post-intervention density $\tilde{g}_{0, A}(A \mid W)$ on
$P$ (i.e., the third term for \textit{learning} a stochastic intervention) may
be dropped. In this case, the EIF may be expressed
\begin{align*}
  D^{\star}(O; P) =&
    \left(Y \frac{\tilde{g}_{A}(A \mid W)}{g_{A}(A \mid W)} -
      \Psi(P)\right)\\ &- \left(\overline{Q}_{Y}(A,W)
      \frac{\tilde{g}_{A}(A \mid W)}{g_{A}(A \mid W)} -
    \int_{\mathcal{A}} \overline{Q}_{Y}(a,W)
      \tilde{g}_{A}(a \mid W) d\mu(a) \right) \ .
\end{align*}
With this in mind, we need only segregate the IPW estimating function $U(g_A,
\psi)$ from the remaining terms in this expression of the EIF $D^{\star}(P)$ in
order to formulate a suitable term to be used as an objective for undersmoothing
over a sequence of conditional density estimators $\{g_{n,A,\lambda}:
\lambda_1, \ldots, \lambda_K\}$. Recalling the general expression for the EIF
$D^{\star}(P)$ as a difference between the IPW estimating function
$U(g_A, \psi)$ and any remaning terms, we denote this expression
$D_{\text{CAR}}$.

By equivalence of these two preceding expressions for the EIF, we now define
$\widetilde{D}_{\text{CAR}} \coloneqq D_{\text{CAR}} - s_g$, where the two terms
are as defined in the general expression of the EIF above. We stress that
$\widetilde{D}_{\text{CAR}}$ is the difference between the projection term in
the case of a known stochastic intervention $\tilde{g}_{A}$ and the score
term $s_g$ arising from the dependence of $\tilde{g}_{A}$ on $P$. By
re-expressing $s_g$, we obtain the following expression for
$\widetilde{D}_{\text{CAR}}$:
\begin{align*}
  \widetilde{D}_{\text{CAR}}(O; P) =&
    \underbrace{\left(\overline{Q}_{Y}(A,W)
      \frac{\tilde{g}_{A}(A \mid W)}{g_{A}(A \mid W)} -
      \int_{\mathcal{A}} \overline{Q}_{Y}(a,W)
      \tilde{g}_{A}(a \mid W) d\mu(a)
      \right)}_{\text{score for a \textit{known} stochastic
      intervention ($D_{\text{CAR}}$)}} \\
    &- \underbrace{\left(\overline{Q}_{Y}^{d}(A, W) -
      \int_{\mathcal{A}} \overline{Q}_{Y}^{d}(a,W)
      g_{A}(a \mid W) d\mu(a)
      \right)}_{\text{score for a \textit{learned} stochastic
      intervention ($s_g$)}} \ .
\end{align*}
Under the MTP definition and using Assumption~\ref{main:ass:piece_inv} to apply
the change-of-variable formula (per~\citet{haneuse2013estimation}) to re-express
the first integral expression identically to the second, the two integral
expressions can be made to cancel out. This resolves a significant practical
obstacle to working with $\widetilde{D}_{\text{CAR}}$ as a selection criterion.
We now have a simplified expression for $\widetilde{D}_{\text{CAR}}$:
\begin{equation*}
  \widetilde{D}_{\text{CAR}}(O; P) = \overline{Q}_{Y}(A,W)
    \frac{\tilde{g}_{A}(A \mid W)}{g_{A}(A \mid W)}
    - \overline{Q}_{Y}^{d}(A, W) \ .
\end{equation*}
This projection term $\widetilde{D}_{\text{CAR}}$ may serve as a suitable
criterion for undersmoothing of a non-stabilized IPW
estimator~\eqref{main:eqn:ipw}; however, it does not extend immediately to the
stabilized IPW estimator~\eqref{main:eqn:ipw_stable}. Next, we repeat the
analysis above, abbreviating mathematical manipulations for the sake of
convenience, to derive the equivalent expression for the stabilized IPW
estimator~\eqref{main:eqn:ipw_stable}.

In the case of the stabilized IPW estimator, the form of the IPW estimating
function and its corresponding $D_{\text{CAR}}$ projection term take slightly
different forms than they do in the case of the non-stabilized IPW estimator.
Specifically, in this case, we have
\begin{equation*}
  U(O; g_A, \psi) = \left(\frac{\tilde{g}_{A}(A \mid W)}
    {g_{A}(A \mid W)}\right)(Y - \psi),
\end{equation*}
for the IPW estimating function $U(g_A, \psi)$ at a distribution $P \in
\mathcal{M}$. Then, the corresponding term $D_{\text{CAR}}$ is
\begin{align*}
  D_{\text{CAR}}(O; P) =&~\Psi(P) \left(1 - \frac{\tilde{g}_{A}(A \mid W)}
    {g_{A}(A \mid W)} \right) \\ &+
    \left(\overline{Q}_{Y}(A,W)
    \frac{\tilde{g}_{A}(A \mid W)}{g_{A}(A \mid W)}
    - \int_{\mathcal{A}} \overline{Q}_{Y}(a,W)
    \tilde{g}_{A}(a \mid W) d\mu(a) \right) \ .
\end{align*}
Notably, the score term $s_g$ for the dependence of $\tilde{g}_{A}(A \mid W)$
on the distribution $P$ remains unaltered. Then, a term appropriate as a target
for undersmoothing of the stabilized IPW estimator, analogous to that for the
non-stabilized IPW estimator, is
%\begin{align*}
  %\widetilde{D}_{\text{CAR}}(P) =&~
    %\underbrace{\Psi(P) \left(1 - \frac{g_{\cdot,A}(d^{-1}(A,W) \mid W)}
      %{g_{\cdot,A}(A \mid W)}\right) + \left(\overline{Q}_{\cdot,Y}(A,W)
      %\frac{g_{\cdot,A}(d^{-1}(A,W) \mid W)}{g_{\cdot,A}(A \mid W)} -
    %\int_{\mathcal{A}} \overline{Q}_{\cdot,Y}(A,W)
      %g_{\cdot,A}(d^{-1}(A,W) \mid W) d\mu(A) \right)}_{\text{$D_{\text{CAR}}$
      %for a \textit{known} stochastic intervention}} \\
    %&- \underbrace{\left(\overline{Q}_{\cdot,Y}(d(A,W), W) -
      %\int_{\mathcal{A}} \overline{Q}_{\cdot,Y}(d(A,W),W)
      %g_{\cdot,A}(A \mid W) d\mu(A) \right)}_{\text{$s_g$, the score term for
      %a \textit{learned} stochastic intervention}} \ .
%\end{align*}
\begin{align*}
  \widetilde{D}_{\text{CAR}}(O; P) =&~\Psi(P)
    \left(1 - \frac{\tilde{g}_{A}(A \mid W)}{g_{A}(A \mid W)}
    \right) \\ &+ \left(\overline{Q}_{Y}(A,W)
    \frac{\tilde{g}_{A}(A \mid W)}{g_{A}(A \mid W)} -
    \overline{Q}_{Y}^{d}(A, W) \right) \ .
\end{align*}
and may equivalently be re-expressed
\begin{align*}
  \widetilde{D}_{\text{CAR}}(O; P) =&~\Psi(P)
    \left(\frac{g_{A}(A \mid W) - \tilde{g}_{A}(A \mid W)}
    {g_{A}(A \mid W)} \right) \\ &+ \left(\overline{Q}_{Y}(A,W)
    \frac{\tilde{g}_{A}(A \mid W)} {g_{A}(A \mid W)} -
    \overline{Q}_{Y}^{d}(A, W) \right) \ .
\end{align*}
As per the main text, this projection of the EIF onto
$\mathcal{F}_{\text{CAR}}$, the space of functions of $A$ that are mean-zero
conditional on $W$ can be used in formulating a targeted undersmoothing
criterion for selecting asymptotically efficient IPW estimators that solve the
EIF estimating equation as per~\eqref{main:eqn:dcar_crit}. Any IPW estimator
sufficiently minimizing $\lvert P_n D_{\text{CAR}}(g_{n,A,\lambda},
\overline{Q}_{n,Y}) \rvert$ is necessarily a solution of $\lvert P_n
D^{\star}(g_{n,A,\lambda}, \overline{Q}_{n,Y}) \rvert \approx 0$, given that all
IPW estimators are constructed as solutions to $\lvert P_n U(g_A, \psi) \rvert
\approx 0$.

\section{Second-order remainder of the efficient influence
function}\label{r2_eif}

To formally characterize the exact second-order remainder $R_2$ of the EIF
expansion~\eqref{main:eqn:r2_expansion}, as presented in
Lemma~\ref{main:lemma:R2}, we consider the EIF expansion at $P, P_0 \in \M$:
\begin{equation*}
  R_2(g_{n,A}, \overline{Q}_{n,Y}, g_{0,A}, \overline{Q}_{0,Y}) =
    \Psi_{\delta}(P) - \Psi_{\delta}(P_0) + P_0 D^{\star}(P),
\end{equation*}
where, per the discussion in Section~\ref{shift_dcar}, the last term
$P_0 D^{\star}(P)$ admits the representation
\begin{align}
  P_0 D^{\star}(O; P) =&~P_0 \overline{Q}_{Y}(A, W)
    \frac{\tilde{g}_{A}(A \mid W)}{g_{A}(A \mid W)} -
    \Psi_{\delta}(P) \label{eqn:r2_exp_p1} \\
    &- P_0 \overline{Q}_{0,Y}(A, W)
    \frac{\tilde{g}_{A}(A \mid W)}{g_{A}(A \mid W)}
    \nonumber \\
    &+ P_{0,W} \mathop{\mathlarger{\int}}\limits_{\mathcal{A}}
    \overline{Q}_{Y}(a,W) \tilde{g}_{A}(a \mid W) d\mu(a)
    \nonumber \\
    =&~ P_0 (\overline{Q}_{Y}(A, W) - \overline{Q}_{0,Y}(A, W))
    \frac{\tilde{g}_{A}(A \mid W)}{g_{A}(A \mid W)}
    g_{0,A}(A \mid W) \label{eqn:r2_exp_p2} \\
    &+ P_{0,W} \mathop{\mathlarger{\int}}\limits_{\mathcal{A}}
    \overline{Q}_{Y}(a,W) \tilde{g}_{A}(a \mid W) d\mu(a)
    - \Psi_{\delta}(P) \nonumber \\
    =&~ P_0 (\overline{Q}_{Y}(A, W) - \overline{Q}_{0,Y}(A, W))
    \left(\frac{g_{0,A}(A \mid W)}{g_{A}(A \mid W)} - 1\right)
    \tilde{g}_{A}(A \mid W) \label{eqn:r2_exp_p3} \\
    &+ P_0(\overline{Q}_{Y}(A, W) - \overline{Q}_{0,Y}(A, W))
    \tilde{g}_{A}(A \mid W) \nonumber
\end{align}
Throughout the manipulations in the expressions above, the rearrangements
between~\eqref{eqn:r2_exp_p1} and~\eqref{eqn:r2_exp_p2} are straightfoward,
while the final equivalence~\eqref{eqn:r2_exp_p3} is justified by use of the
true distribution of $W$, under which the equivalence $P_0 \int_{\mathcal{A}}
\overline{Q}_{Y}(a,W) \tilde{g}_{A}(a \mid W) d\mu(a) - \Psi_{\delta}(P) = 0$
holds. In order to obtain an exact expression for the second-order remainder
term, one need only combine~\eqref{eqn:r2_exp_p3} with the difference term
$\Psi_{\delta}(P) - \Psi_{\delta}(P_0)$, yielding $R_2(g_{n,A},
\overline{Q}_{n,Y}, g_{0,A}, \overline{Q}_{0,Y})$, per the definition above.
Combining these expressions, and after mathematical simplifications omitted for
brevity, one arrives immediately at the form of the remainder given in
Lemma~\ref{main:lemma:R2}:
\begin{align*}
  R_2(&g_{A}, \overline{Q}_{Y}, g_{0,A}, \overline{Q}_{0,Y}) = \\
  &\mathop{\mathlarger{\int}}\limits_{\mathcal{W}}
  \mathop{\mathlarger{\int}}\limits_{\mathcal{A}}
  \left(\overline{Q}_{Y}(a, w) - \overline{Q}_{0,Y}(a, w)\right)
  \left(\frac{g_{0,A}(a \mid w) - g_{A}(a \mid w)}{g_{A}(a \mid w)}
  \right) \tilde{g}_{A}(a \mid w) d\mu(a) d\nu(w)\ .
\end{align*}
This form of the second-order remainder provides insights into the mechanism by
which undersmoothing of an estimator of $g_{0,A}(A \mid W)$ can endow an
estimator $\Psi_{\delta}(P)$ of $\Psi_{\delta}(P_0)$ with desirable asymptotic
efficiency properties.

In the \href{paper}{main manuscript}, shortly after presenting
Lemma~\ref{main:lemma:R2}, it is noted that this remainder term may be
characterized in terms of $D_{\text{CAR}}$, that is, $R_2(g_{A},
\overline{Q}_{Y}, g_{0,A}, \overline{Q}_{0,Y}) \equiv P_0 D_{\text{CAR}}(g_{A},
\overline{Q}_{Y} - \overline{Q}_{0,Y})$. This expression for the second-order
remainder is based on re-expressing the EIF using the characterization
introduced in Lemma~\ref{main:lemma:dcar}, which explicitly states that
$D^{\star}(P) = U(g_A, \psi) - D_{\text{CAR}}(P)$. Straightforward manipulation
of this expression of the EIF quickly yields,
\begin{align*}
P_0 D^{\star}&(g_{A}, \overline{Q}_{Y}) -
P_0 D^{\star}(g_{A}, \overline{Q}_{0,Y}) = \\
&P_0 \left(\frac{\tilde{g}_{A}}{g_{A}} Y
\right) - \Psi_{\delta}(P)
- P_0 \left(\frac{\tilde{g}_{A}}{g_{A}} Y
\right) + \Psi_{\delta}(P_0) \\
&- P_0 D_{\text{CAR}}(\overline{Q}_{Y}, g_{A})
+ P_0 D_{\text{CAR}}(\overline{Q}_{0,Y}, g_{A}) \\
=&~\Psi_{\delta}(P_0) - \Psi_{\delta}(P) +
P_0 D_{\text{CAR}}(g_{A}, \overline{Q}_{Y} - \overline{Q}_{0,Y}) \ .
\end{align*}
With this equivalence established, it is now easier to notice that the
second-order remainder is itself only a score of the treatment mechanism; thus,
undersmoothing of an estimator of the treatment mechanism would be expected to
solve $R_2$ to some nontrivial degree. Indeed, this expression provides the
insight that undersmoothing of a HAL estimator of $g_{0,A}(A \mid W)$ so as to
include a richer set of basis functions (yielding a more nonparametric
estimator) ought to solve this second-order remainder better. As noted in the
\href{paper}{main manuscript}, $R_2$ itself, unlike $D_{\text{CAR}}$, cannot be
used as an undersmoothing target, since its form relies on the true outcome
mechanism $\overline{Q}_{0,Y}(A,W)$, whose form is never known in practice. Yet,
it may be possible to make use of $R_2$ as a diagnostic for the comparison of
candidate estimators. To this end, in our numerical experiments (presented in
Section~\ref{main:sim} and reexamined in Section~\ref{addl_sim_res}), we
evaluate the performance of our proposed IPW estimators, alongside that of
several competitors, using an approximation of $R_2$ (as well as standard
metrics). To do so, we define $\tilde{R}_{2}(\cdot) \coloneqq P_0
D_{\text{CAR}}(g_{A}, \overline{Q}_{Y} - \overline{Q}_{0,Y})$,
which can be evaluated in any given \textit{simulated} dataset $O_1, \ldots,
O_n$ by estimating the nuisance quantities (e.g., with appropriate HAL
estimators) and computing the differences of these against the known forms of
the nuisance parameters, themselves evaluated on the given dataset. In our
examples, by applying the MTP directly to the given dataset, it is possible to
numerically evaluate the integral expressions (over $\tilde{g}_{A}(A \mid
W)$)) that appear in the definition of the remainder term. Evaluating
$\tilde{R}_2$ across many numerical examples then allows approximation of $R_2$
as $R_2 = P_0 \tilde{R}_2$, where the expectation with respect to $P_0$ is
evaluated by Monte Carlo integration.

\section{Alternative conditional density estimators}\label{alt_dens}

\subsection{Semiparametric location-scale conditional density
estimation}\label{hose_hese_est}

A recently developed family of flexible semiparametric conditional density
estimators takes the general form $\rho(A - \mu(W) / \sigma(W))$, where $\rho$
is a given marginal density function. Conditional density estimation procedures
falling within this framework may be characterized as belonging to
a \textit{conditional location-scale} family, i.e., where $g_{n,A}(A \mid W)
= \rho((A - \mu_n(W)) / \sigma_n(W))$; where we stress that the marginal density
mapping $\rho$ is selected \textit{a priori}, leaving only the relevant moments
$\mu$ and $\sigma$ to be estimated.

Though the restriction to (conditional) location-scale families imposes some
limitations on the form of the target functional, the strategy is made
particularly flexible by its ability to incorporate arbitrary, data-adaptive
regression strategies for the estimation of $\mu(W) = \E[A \mid W]$ and,
optionally, of the conditional variance $\sigma(W) = \E[(A - \mu(W))^2 \mid W]$.
In particular, in settings with limited data, the additional structure imposed
by the assumption of form of the target density functional (i.e., in the
specified kernel function $\rho$) can prove beneficial, when the true density
function admits such a representation. While we stress that this procedure is
surely not a novel contribution of the present work, we have been otherwise
unable to ascertain a formal description of it; thus, we provide such
a formalization in Algorithm~\ref{alg:loc_scale_dens}, which sketches the
construction of conditional density estimators of this family.

\begin{algorithm}[H]
\SetAlgoLined
\KwResult{Estimates of the conditional density of $A$, given $W$.}
\SetKwInOut{Input}{Input}\SetKwInOut{Output}{Output}
\Input{
  \begin{itemize}
    \item[1] An observed data vector of the continuous treatment for $n$
      units: $A$
    \item[2] An observed data vector (or matrix) of the baseline covariates for
      $n$ units: $W$
    \item[3] A kernel function specification to be used to construct the
       density estimate: $\rho$
    \item[4] A candidate regression procedure to estimate the conditional mean
      $\mu(W)$: $f_{\mu}$
    \item[5] A candidate regression procedure to estimate the conditional
    variance $\sigma(W)$: $f_{\sigma}$
  \end{itemize}
}
\BlankLine

Estimate $\E[A \mid W]$, the conditional mean of $A$ given $W$, by applying the
regression estimator $f_{\mu}$, yielding $\hat{\mu}(W)$.

Estimate $\mathbb{V}[A \mid W]$, the conditional variance of $A$ given $W$, by
applying the regression estimator $f_{\sigma}$, yielding $\hat{\sigma}^2(W)$.
Note that this step involves only estimation of the conditional mean $\E[(A
- \hat{\mu}(W))^2 \mid W]$.

Estimate the one-dimensional density of $(A - \hat{\mu}(W))^2
/ \hat{\sigma}^2(W)$, using kernel smoothing to obtain $\hat{\rho}(A)$.

Construct the estimated conditional density $g_{n,A}(A \mid W) = \hat{\rho}((A
- \hat{\mu}(W)) / \hat{\sigma}(W))$.
\BlankLine

\Output{
  $g_{n,A}$, an estimate of the generalized propensity score.
}
\caption{Location-scale conditional density estimation}
\label{alg:loc_scale_dens}
\end{algorithm}

The sketch of the algorithm for constructing estimators $g_{n,A}(a \mid w)$ of
the generalized propensity score may take two forms, which diverge at the second
step above. Firstly, one may elect to estimate only the conditional mean
$\mu(W)$ via a regression technique, leaving the variance to be taken as
constant (estimated simply as the marginal mean of $\E[(A - \hat{\mu}(W))^2]$).
Estimators of this form may be described as having \textit{homoscedastic error},
based on the variance assumption made. Alternatively, one may additionally
estimate the conditional variance $\sigma^2(W)$ via the residuals of the
estimated conditional mean, that is, estimating instead the conditional mean
$\E[(A - \hat{\mu}(W))^2 \mid W]$.

While the regression procedures $f_{\mu}$ and $f_{\sigma}$ used to estimate the
conditional mean $\mu(W)$ and the conditional variance $\sigma^2(W)$,
respectively, may be arbitrary, with candidates including, for example,
%random forests~\citep{breiman2001random}, spline
%regression~\citep[e.g.,][]{stone1994use}, or
regression ensembles~\citep{breiman1996stacked,vdl2007super}, we recommend the
use of HAL regression~\citep{benkeser2016highly,vdl2017generally}, as its use
will ensure an enhanced rate of convergence~\citep{bibaut2019fast} of the
estimator $g_{n,A}$ to its target $g_{0,A}$. The data-adaptive nature of HAL
regression affords a degree of flexibility that ought to limit opportunities for
model misspecification to compromise estimation of $g_{0,A}$; morever, their
improved convergence rate will help to facilitate the construction of
asymptotically linear and efficient estimators of $\psi_{0, \delta}$.
%In Section~\ref{sim}, we compare doubly robust estimators based on these
%estimators of $g_{n,A}$ under both flexible (HAL) and parametric (GLM)
%estimators of the conditional moments.

\subsection{Proportional hazards highly adaptive lasso conditional density
estimation}\label{cox_hal}

Note that a conditional density may be written in terms of the hazard function
and survival function as follows
\begin{equation}\label{eqn:cond_dens}
  g_A(a \mid w) = \lambda(a \mid w) \exp\left(-\int_0^a \lambda(s \mid w)
    ds\right),
\end{equation}
where $\lambda(a \mid w)$ is the conditional hazard function and the survival
function is $\exp(-\int_0^a \lambda(s \mid w) ds$, i.e., a re-scaled cumulative
hazard. Using the highly adaptive lasso, the conditional hazard may be estimated
based on a Cox proportional hazard model:
\begin{equation}\label{eqn:cox_hal}
  \lambda(a \mid w) = \lambda_0(a) \exp\left(\sum_j \beta_j \phi_j\right),
\end{equation}
where any given $\phi_j = \mathbb{I}(a > c_j)$ for some cutoff $c_j$,
essentially a knot point when spline basis functions are usd. Assuming that the
quantity in equation~\eqref{eqn:cox_hal} can be estimated using open source
software, as available in the \texttt{hal9001}~\citep{hejazi2020hal9001-joss,
coyle2021hal9001-rpkg} and \texttt{glmnet}~\citep{friedman2009glmnet} \texttt{R}
packages. In the sequel, we focus on estimation of the quantity $\Lambda(a \mid
w) = \int_0^a \lambda(s \mid w) ds$, the cumulative hazard, assuming access to
the quantity $\lambda(a \mid w)$ since its estimation is straightforward in
survival analysis.

Note that the cumulative hazard function is merely the primitive of the
conditional hazard function $\lambda(a \mid w)$; thus, we have
\begin{align*}
  \Lambda(a \mid w) &= \int_0^a \lambda(s \mid w) ds\\
              &= \int_{s=0}^a \lambda_0(s) \exp\left(\sum_j \beta_j \phi_j(s,w)
                  \right) ds\\
              &=\sum_{l=1}^L \lambda_0(s) \exp\left(\sum_j \beta_j
                  \phi_j(a_l, w) (a_{l+1} - a_l)\right),
\end{align*}
where $a_l \leq a$ and this final quantity is estimable via standard techniques
for numerical integration. With an estimate of $\Lambda(a \mid w)$ in hand, an
estimator of the conditional density may be constructed by the relation given in
equation~\eqref{eqn:cond_dens}:
\begin{equation*}
  g_A(a \mid w) = \lambda(a \mid w) \exp(-\Lambda(a \mid w)) \ .
\end{equation*}
Although it is not without assumptions --- most notably, that of proportional
hazards --- such an estimator of the conditional density is more flexible than
its entirely parametric counterparts, especially when nonparametric regression
(e.g., HAL) is used for estimation of the conditional hazard function. A variant
of this strategy has been explored and successfully utilized
by~\citet[][Appendix B and C]{rytgaard2022continuous} in the context of
conditional density estimation in longitudinal studies using an intensity-based
representation of the hazard based on the Poisson likelihood.

%%%%%%%%%%%%%%%%%%%%%%%%%%%%%%%%%%%%%%%%%%%%%%%%%%%%%%%%%%%%%%%%%%%%%%%%%%%%%%%

\section{Doubly robust estimation}\label{dr_est}

Two popular frameworks for efficient estimation, both taking advantage of the
EIF estimating equation, include one-step
estimation~\citep{pfanzagl1985contributions,bickel1993efficient} and targeted
minimum loss (TML) estimation~\citep{vdl2006targeted, vdl2011targeted}.
Importantly, both the one-step and TML estimators are \textit{doubly robust}
(DR), a property which affords two important conveniences. Firstly, both
estimators are consistent for $\psi_{0,\delta}$ when either of the initial
nuisance estimates of $g_{n,A}$ and $\overline{Q}_{n,Y}$ are consistent for
their respective nuisance functionals. While the protection afforded by this
form of robutness is practically useful, it carries with it a disadvantage:
these estimators can only achieve asymptotic efficiency when \textit{both}
initial nuisance estimates are consistent. Secondly, as a consequence of their
explicit construction based on the EIF, both readily accommodate the use of
flexible, data-adaptive estimation strategies for the initial estimation of
nuisance parameters. This latter property provides these estimators with
a distinct advantage over their direct and IPW estimator counterparts: two
opportunities to avoid misspecification bias.
%arising from reliance upon restrictive (e.g., parametric) modeling strategies.

Invoking either estimation strategy proceeds first by constructing initial
estimators, $g_{n,A}$ and $\overline{Q}_{n,Y}$, of the conditional treatment
density $g_{0,A}$ and the outcome mechanism $\overline{Q}_{0,Y}$. These two DR
frameworks diverge at their second stage, each of which implement different
forms of bias correction. Within the one-step framework, this amounts to
updating the initial estimate $\psi_{n,\delta}$ --- itself obtained simply by
``plugging in'' the initial estimate of the outcome mechanism
$\overline{Q}_{n,Y}$ into the direct estimator~\eqref{main:eqn:plugin} --- by
adding to it the empirical mean of the EIF evaluated at the initial nuisance
estimates $\{g_{n,A}, \overline{Q}_{n,Y}\}$. Using the EIF $D^{\star}_{n,i}$
evaluated at the nuisance estimates $\{g_{n,A}, \overline{Q}_{n,Y}\}$, the
one-step estimator is
%\begin{equation}\label{eqn:one_step}
$\psi_{n,\delta}^{+} \coloneqq \psi_{n,\delta} + n^{-1} \sum_{i=1}^n
D^{\star}_{n,i}.$
%\end{equation}
The one-step estimator $\psi_{n,\delta}^{+}$ of $\psi_{0,\delta}$ is both
asymptotically linear (on account of the bias correction step) and capable of
asymptotically achieving the nonparametric efficiency bound; moreover, it is
compatible with initial nuisance estimates $\{g_{n,A}, \overline{Q}_{n,Y}\}$
that are constructed from flexible, data-adaptive regression strategies. On
account of its additive bias-correction procedure, the one-step estimator is not
guaranteed to respect the bounds of the parameter space of $\psi_{0,\delta}$.

In the TML estimation framework, a univariate (most often, logistic) parametric
tilting model is used to update the initial estimate $\overline{Q}_{n,Y}$ of the
outcome mechanism in such a way that the EIF estimating equation is solved up to
a theoretically desirable degree. Plugging this updated initial estimate
$\overline{Q}_{n,Y}^{\star}$ into the direct estimator~\eqref{main:eqn:plugin}
yields a \textit{targeted} estimator $\psi_{n,\delta}^{\star}$ of
$\psi_{0,\delta}$. Specifically, the TML estimator may be constructed by
updating the initial estimator $\overline{Q}_{n,Y}$ to a tilted variant
$\overline{Q}_{n,Y}^{\star}$, through a one-dimensional logistic tilting model
of the form $\logit \overline{Q}_{n,Y}^{\star} = \logit \overline{Q}_{n,Y}
+ \epsilon H_n$, in which the initial estimate $\overline{Q}_{n,Y}$ is taken as
an offset (i.e., fixed parameter value known \textit{a priori} to be one) and
only the parameter $\epsilon$ need be estimated. This targeted estimator is then
%\begin{equation}\label{eqn:tmle}
$\psi_{n,\delta}^{\star} \coloneqq \int \overline{Q}_{n,Y}^{\star}(d(a, w;
\delta), w) dQ_{n,AW}(a,w).$
%\end{equation}
Like its one-step estimator counterpart, the TML estimator
$\psi_{n,\delta}^{\star}$ is both asymptotically linear (on account of the bias
correction afforded by the tilting step) and capable of asymptotically achieving
the nonparametric efficiency bound. It is also compatible with initial nuisance
estimates $\{g_{n,A}, \overline{Q}_{n,Y}\}$ that are constructed from flexible,
data-adaptive regression strategies, and, unlike the one-step estimator, it is
a plug-in estimator, incapable of lying outside the bounds of the parameter
space of $\psi_{0, \delta}$. Additionally, numerical studies have shown that the
TML estimator generally displays better finite-sample performance than the
one-step estimator~\citep{vdl2011targeted}. Accordingly, in the numerical
experiments presented in Section~\ref{main:sim} of the manuscript, we use a TML
estimator under differing nuisance parameter configurations. Both one-step and
TML estimators of $\psi_{0,\delta}$ are implemented in the open source
\texttt{txshift} \texttt{R} package~\citep{hejazi2020txshift-joss,
hejazi2022txshift-rpkg}, freely available on the Comprehensive \texttt{R}
Archive Network.

%%%%%%%%%%%%%%%%%%%%%%%%%%%%%%%%%%%%%%%%%%%%%%%%%%%%%%%%%%%%%%%%%%%%%%%%%%%%%%%
\section{Additional numerical results}\label{addl_sim_res}

Here, we consider the performance of the same set of six IPW and four DR
estimators across the two DGPs discussed in the \href{paper}{main manuscript},
though we assess performance in terms of three alternative metrics. These three
metrics are the empirical coverage of oracle confidence intervals (CIs), which
use a Monte Carlo variance estimate to avoid issues introduced by unstable
variance estimation; the empirical coverage of Wald-style confidence intervals,
which use the empirical variance of the \textit{estimated} EIF and so may be
prone to issues of poor variance estimation; and the magnitude of an
approximation of the second-order remainder $R_2$ associated with each of the
estimator variants. The manner in which this second-order remainder term is
approximated is briefly discussed in Section~\ref{r2_eif}. Since this
approximation of the remainder term $R_2$ is prone to a small degree of
instability, we compute the metric using a trimmed mean that removes the lowest
and highest 5\% of its realizations across the $300$ simulations. We note that
these metrics are similar to those considered in Section~\ref{main:sim} of the
\href{paper}{main manuscript}: performance in terms of the coverage of oracle
CIs is a measure of bias; that in terms of the coverage of Wald-style CIs is
a measure of efficiency (in that it accounts for both bias and the quality of
variance estimation); and that in terms of the magnitude of $R_2$ is a measure
of higher-order efficiency, that is, how well the estimators achieve efficiency
\textit{beyond} the first-order EIF approximation.

\subsection{Simulation \#1: Treatment mechanism with equal mean and
  variance}\label{hese_sim_poisson_res_addl}

\begin{figure}[H]
  \centering
  \includegraphics[scale=0.28]{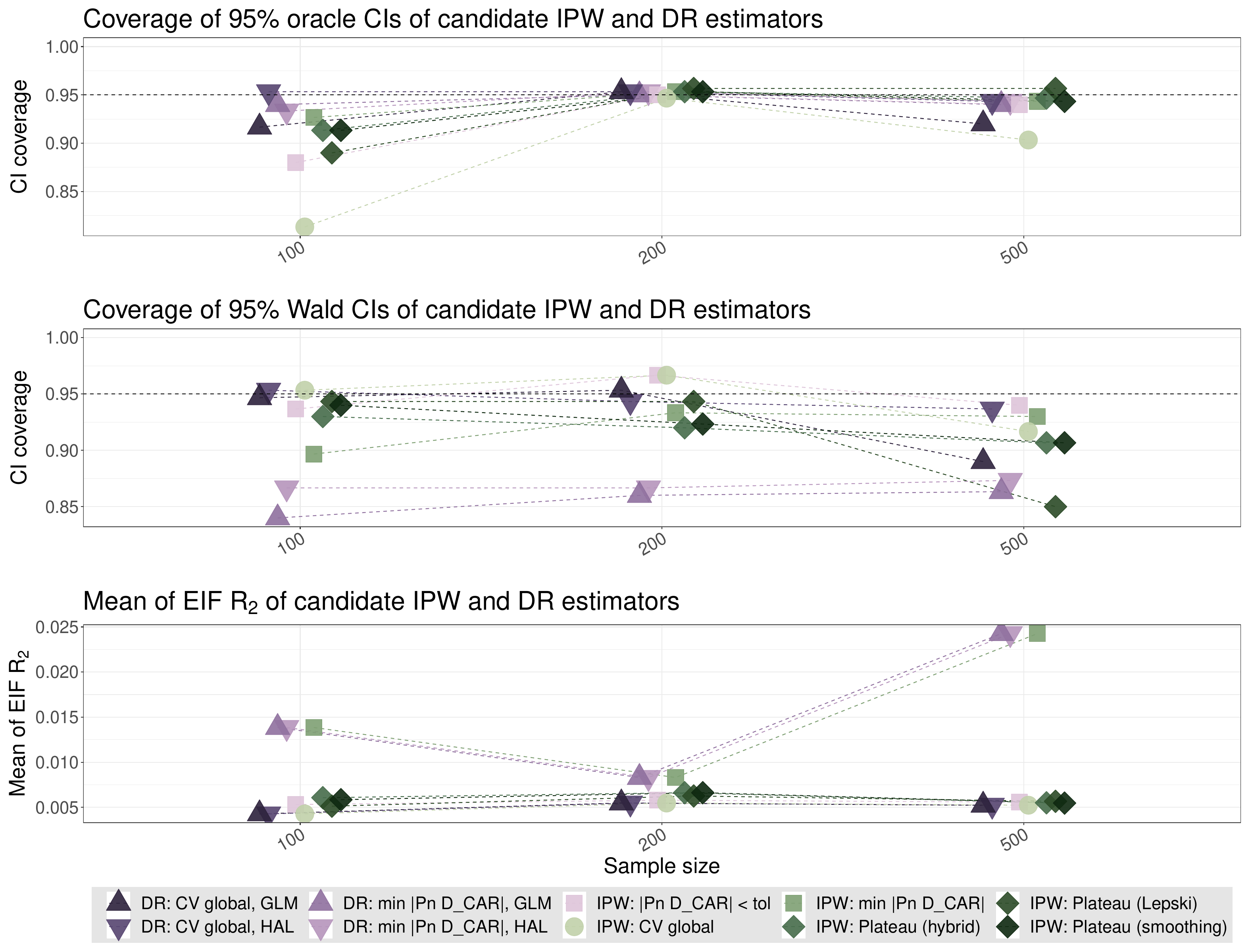}
  \caption{Alternative numerical comparisons of our nonparametric IPW
    estimators and DR estimators.}
  \label{fig:dgp2a_npipw_addl}
\end{figure}

Using the same DGP as considered in Section~\ref{main:hese_sim_poisson}, we
evaluate each of the estimator variants, in terms of the three alternative
metrics discussed above, across $300$ simulation experiments. Their performance
is summarized in Figure~\ref{fig:dgp2a_npipw_addl}.

Examination of Figure~\ref{fig:dgp2a_npipw_addl} quickly reveals that all but
one of the estimators considered perform well in terms of the coverage of oracle
CIs, indicating negligible bias for the vast majority of estimators. This
assessment is generally in agreement with corresponding results on bias
appearing in Figure~\ref{main:fig:dgp2a_npipw}. Notably, the IPW estimator based
on ``global'' CV selection shows poor coverage (20\% lower than the nominal
rate) at $n = 100$ and degrading coverage from $n = 200$ (at the nominal rate)
and $n = 500$ (5\% lower than the nominal rate); again, this is consistent with
conclusions drawn based on other metrics. The coverage of Wald-style CIs reveals
that the IPW estimator variants based on targeted undersmoothing perform
comparably to the best variant among the DR estimators. The IPW estimators based
on agnostic undersmoothing achieve coverage near the nominal rate, but show very
limited coverage loss (5\% lower than the nominal rate) by $n = 500$; moreover,
the IPW estimator using the agnostic criterion~\eqref{main:eqn:plateau_var_cond}
shows significant loss of coverage (10\% lower than the nominal rate) by $n
= 500$, suggesting instability of this selector, possibly on account of poor
standard error estimation. The IPW estimator based on the ``global'' CV selector
appears to cover near the nominal rate, though this is likely attributable to
the manner in which this selector priorities variance estimation over debiasing.
Among the DR estimators, those basing the treatment mechanism estimate on the
``global'' CV selector achieve desirable coverage at $n = 100$ and $n = 200$,
but the variant using a misspecified GLM to estimate the outcome mechanism shows
degrading coverage by $n = 500$. Those DR estimators using
criterion~\eqref{main:eqn:dcar_crit} for the treatment mechanism estimator
exhibit poorer coverage at $n = 100$ and $n = 200$ ($\approx$7\% lower than the
nominal rate) but but appear to be steadily improving with growing sample size.
Turning now to the metric approximating the second-order remainder $R_2$, most
of the estimator variants display comparably good performance uniformly across
the three sample sizes; however, the DR estimators and the undersmoothed IPW
estimator that make use of criterion~\eqref{main:eqn:dcar_crit} for treatment
mechanism estimation result in an inflated second-order remainder relative to
their competitors (by a factor of three at $n = 100$ and by a factor of five at
$n = 500$). This suggests that absolute minimization of $\lvert P_n
\widetilde{D}_{\text{CAR}}(g_{n,A,\lambda}, \overline{Q}_{n,Y}) \rvert$ carries
with it a tradeoff as regards stability of the selected treatment mechanism
estimator, as related IPW estimators exhibit low bias and reasonable efficiency
but may be doing so by compromising higher-order efficiency. This is not
necessarily at odds with theoretical expectations. Overall, the picture painted
by examining our proposed IPW estimators and standard DR competitors along this
alternative set of metrics speaks to the reliable performance of our
undersmoothed IPW estimators, in terms of inferential bias (captured by the
coverage of CIs) and higher-order efficiency criteria.

\begin{figure}[H]
  \centering
  \includegraphics[scale=0.32]{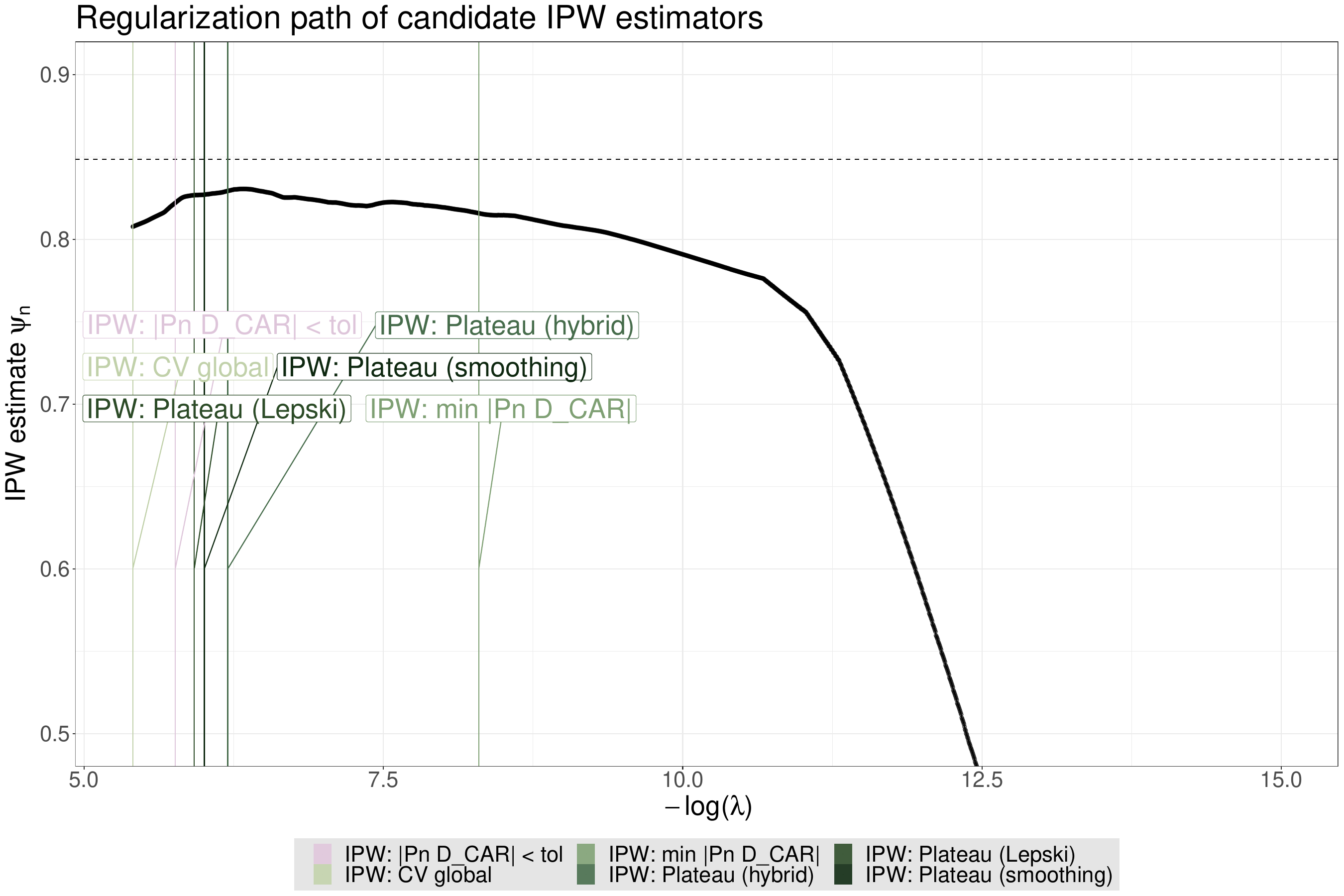}
  \caption{Solution path in $\lambda$ of nonparametric IPW estimators in a
    single numerical example.}
  \label{fig:dgp2a_npipw_solpath}
\end{figure}

In order to gain a degree of intuition regarding the choices of treatment
mechanism estimator made by each of the selectors, we examine, in a single
numerical example with sample size $n = 100$, all of the IPW estimator
candidates along the regularization path defined by $\lambda$. Initially, this
regularization path is set to include $3000$ preset values of $\lambda$ (fixed
across all of our numerical experiments), though the $\lambda_n$ chosen by the
``global'' CV selector is used as the starting point for the
undersmoothing-based selectors. Figure~\ref{fig:dgp2a_npipw_solpath} illustrates
the trajectory taken by the IPW estimators along this grid in $\lambda$.

In Figure~\ref{fig:dgp2a_npipw_solpath}, the six candidate IPW estimators
generated by each of the selectors are displayed, with the choice made by the
``global'' CV selector being the first in the solution path; all values along
the grid corresponding to stricter regularization than this are truncated. The
true value of $\psi_{0,\delta}$, approximated in a very large sample, is
indicated by horizontal dashed line. While this is only a single numerical
example, the IPW estimators chosen by the various selectors may be somewhat
indicative of the degree of regularization preferred by each, and one may
generally expect that the targeted undersmoothing selector based on
criterion~\eqref{main:eqn:dcar_crit} will prefer far less extreme levels of
regularization than the others. Notably, as the degree of regularization is
relaxed too much, the solution path of the IPW estimators diverges (after
$-\log(\lambda) \approx 10$) from the true value $\psi_{0,\delta}$ in an
irrecoverable manner.

\subsection{Simulation \#2: Treatment mechanism with unequal mean and
  variance}\label{hese_sim_negbinom_res_addl}

\begin{figure}[H]
  \centering
  \includegraphics[scale=0.28]{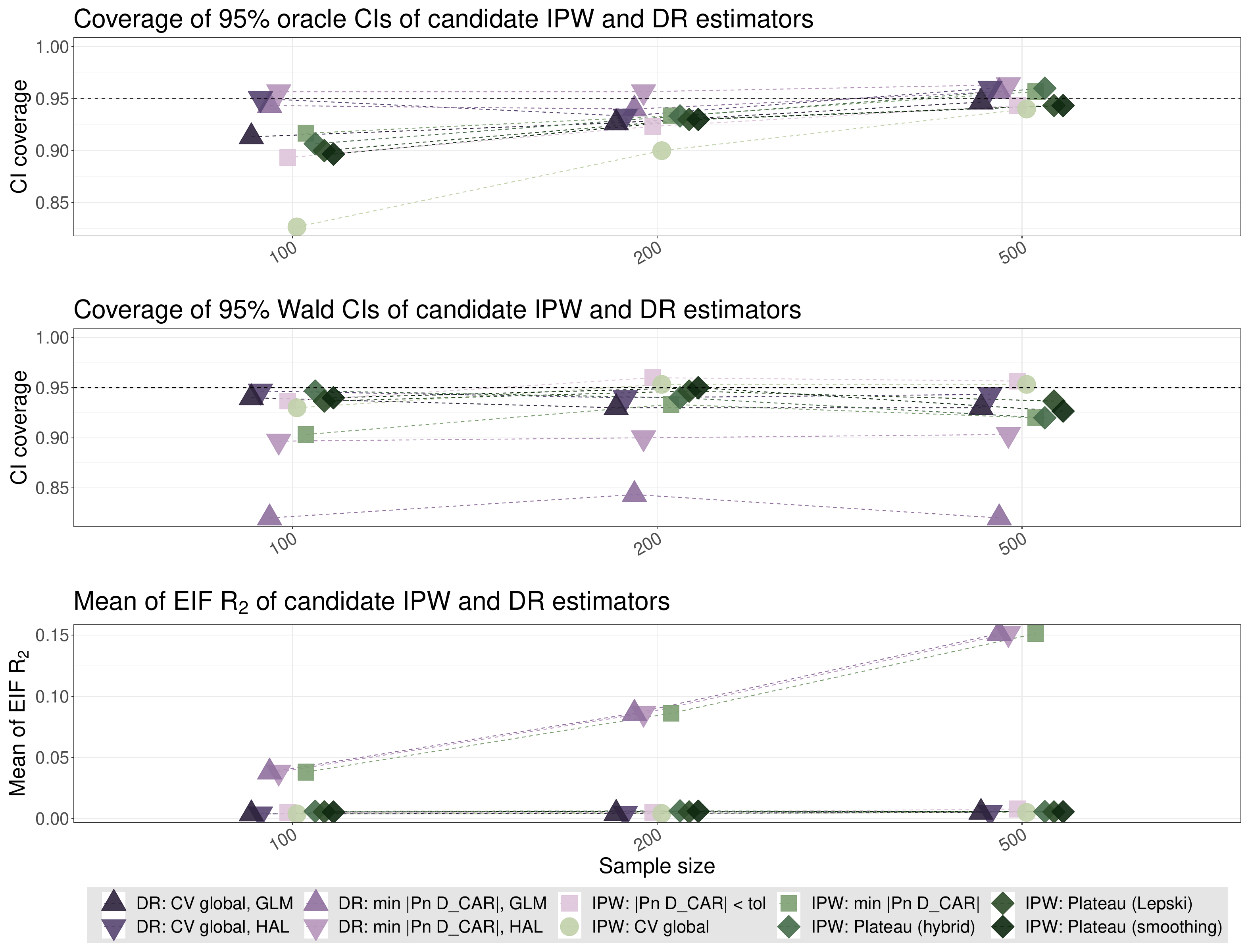}
  \caption{Alternative numerical comparisons of our nonparametric IPW
    estimators and DR estimators.}
  \label{fig:dgp2b_npipw_addl}
\end{figure}

Using the same DGP as considered in Section~\ref{main:hese_sim_negbinom}, we
evaluate each of the estimator variants, in terms of the three alternative
metrics discussed above, across $300$ simulation experiments. Their performance
is summarized in Figure~\ref{fig:dgp2b_npipw_addl}.

Figure~\ref{fig:dgp2b_npipw_addl} reveals fairly good performance, by $n = 200$
or $n = 500$, from all of the estimator variants considered. Examining first the
coverage of oracle CIs, all of the DR estimators achieve coverage at the nominal
rate, irrespective of sample size, indicating unbiasedness of these estimators.
The IPW estimators achieve comparable performance by $n = 500$, though they fare
slightly worse at $n = 100$. In terms of the coverage of Wald-style CIs, all of
the IPW estimators, and half of the DR estimators, achieve coverage at or very
near the nominal rate uniformly across the three sample sizes. The two DR
estimators that base their treatment mechanism estimator on
criterion~\eqref{main:eqn:dcar_crit} display poorer coverage, worse when the
outcome mechanism estimator is misspecified but (predictably) better when the
outcome mechanism estimator is modeled flexibly. Notably, all of the
undersmoothing-based IPW estimators provide reliable coverage (at or near the
nominal rate), indicating their simultaneous achieving negligible bias and
accurate variance estimation. Cursory examination of the approximate
second-order remainder $R_2$ reveals that all but three of the estimator
variants succeed in successfully minimizing this metric. All of the IPW
estimators based on agnostic undersmoothing fall in this majority. As with the
preceding scenario, the DR estimators and the undersmoothed IPW estimator that
make use of criterion~\eqref{main:eqn:dcar_crit} for treatment mechanism
estimation struggle to approppriately minimize this metric, diverging with
growing sample size. This provides further confirmation of the notion that
absolute minimization of $\lvert P_n
\widetilde{D}_{\text{CAR}}(g_{n,A,\lambda}, \overline{Q}_{n,Y}) \rvert$ induces
a compromise in terms of higher-order efficiency, as measured by this remainder
term in the EIF expansion. This may prove a fruitful avenue for future
theoretical and numerical investigation. Overall, the conclusion appears
well-supported that our proposed IPW estimators fare comparably (well) to
standard DR competitors in terms of both inferential bias and higher-order
contributions to asymptotic efficiency.

\begin{figure}[H]
  \centering
  \includegraphics[scale=0.32]{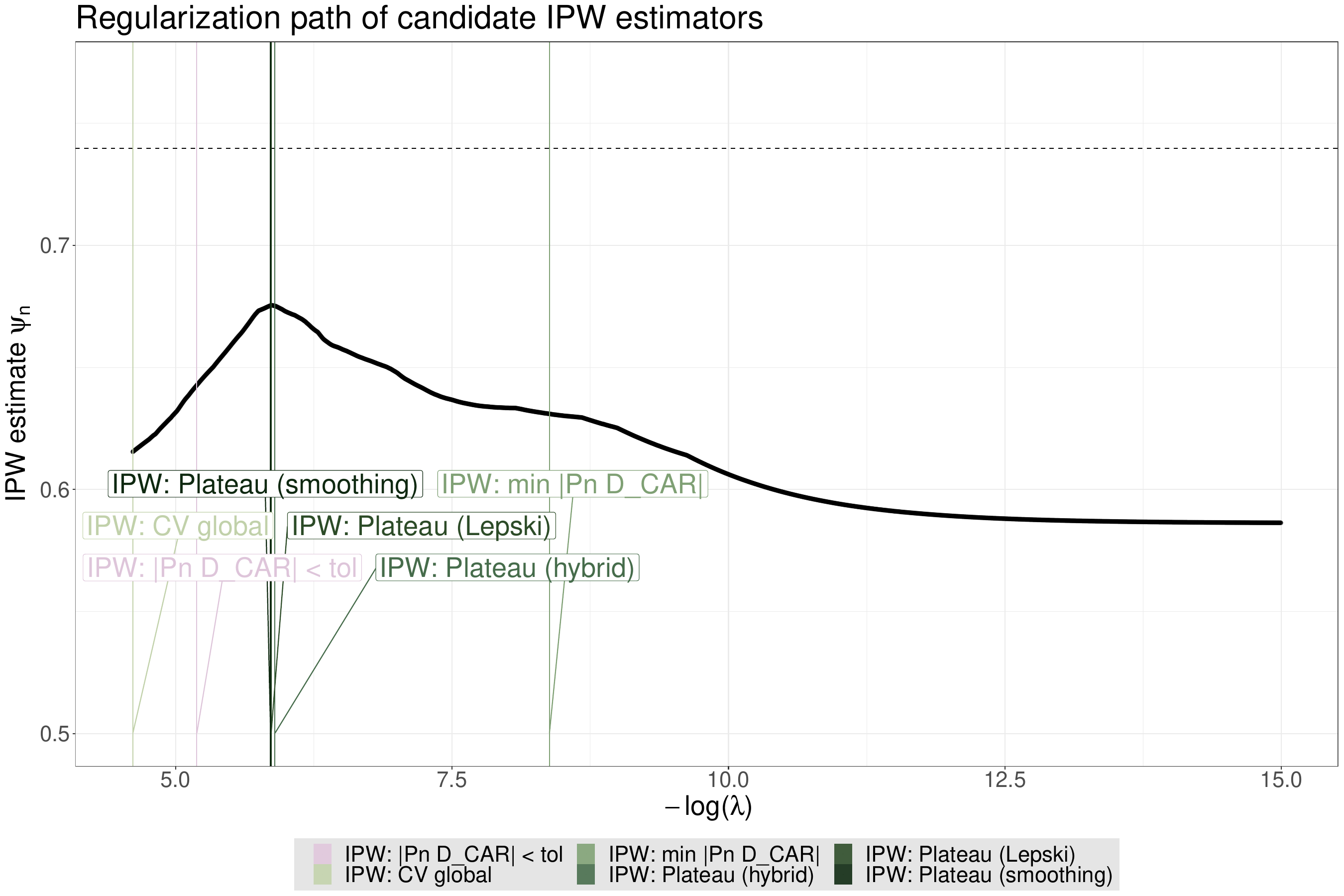}
  \caption{Solution path in $\lambda$ of nonparametric IPW estimators in a
    single numerical example.}
  \label{fig:dgp2b_npipw_solpath}
\end{figure}

As with the previous scenario, we examine a single numerical example at sample
size $n = 100$ to develop some degree of intuition on the choices of treatment
mechanism estimator made by each of the selection criteria. Here, we visualize
the trajectory of the IPW estimators along the regularization path of $3000$
preset values of $\lambda$ (fixed across all of our numerical experiments),
limiting the estimators displayed by truncating the regularization grid so as to
begin at the $\lambda_n$ chosen by the ``global'' CV selector.
Figure~\ref{fig:dgp2b_npipw_solpath} depicts the trajectory taken by the IPW
estimators across this regularization grid.

In Figure~\ref{fig:dgp2b_npipw_solpath}, the six candidate IPW estimators
generated by each of the selectors are displayed, and the ``global'' CV
selector's choice is the first in the solution path; all stricter regularization
values are truncated. The true value of $\psi_{0,\delta}$, approximated in
a very large sample, is indicated by horizontal dashed line. Although this
illustration arises only from a single numerical example, the IPW estimators
chosen by the various undersmoothing selectors can be taken as somewhat
representative of the the degree of regularization preferred by each. For
example, one notices, as with the previous example depicted in
Figure~\ref{fig:dgp2a_npipw_solpath}, that the targeted undersmoothing selector
based on criterion~\eqref{main:eqn:dcar_crit} prefers the most relaxed (least
extreme) degree of regularization; furthermore, one should note that the
agnostic undersmoothing selectors appear to all reliably choose the peak along
the trajectory of IPW estimators, which corresponds to the point estimate
closest to the true value of $\psi_{0,\delta}$, illustrating the debiasing
induced by these undersmoothing selectors.

\newpage
\bibliography{refs}

%% file: _preamble.tex
\usepackage{graphicx}
\usepackage{float}
\usepackage{lmodern}
\usepackage{booktabs}
\usepackage{amsmath,amsfonts,bbm,bm,mathtools,relsize,exscale}
\usepackage{dsfont}
\usepackage[OT1]{fontenc}
\usepackage[utf8]{inputenc}
\usepackage[linesnumbered,ruled,vlined]{algorithm2e}
\usepackage[inline,shortlabels]{enumitem}
\graphicspath{ {figs/} }

\usepackage{refcount,nameref,zref-xr,zref-user} % nameref before zref-xr
\zxrsetup{toltxlabel} % allows use of the \ref command as usual
\usepackage{url}

\let\href\undefined
\usepackage[colorlinks,citecolor=blue,urlcolor=blue]{hyperref}

\makeatletter%
\@ifclassloaded{biometrika}{%
  \let\latexarabic\arabic
  \let\latexdocument\document
  \let\latexenddocument\enddocument

  \RequirePackage[thmmarks]{ntheorem}
  \makeatletter
  \renewtheoremstyle{plain}
    {\item[\hskip\labelsep \theorem@headerfont ##1\ \textup{##2}\theorem@separator]}
    {\item[\hskip\labelsep \theorem@headerfont ##1\ \textup{##2}\ (##3)\theorem@separator]}
  \makeatother

  \makeatletter
  \renewcommand{\algocf@captiontext}[2]{#1\algocf@typo. \AlCapFnt{}#2}
  \def\@algocf@capt@plain{top}
  \renewcommand{\algocf@makecaption}[2]{%
    \addtolength{\hsize}{\algomargin}%
    \sbox\@tempboxa{\algocf@captiontext{#1}{#2}}%
    \ifdim\wd\@tempboxa >\hsize%     % if caption is longer than a line
      \hskip .5\algomargin%
      \parbox[t]{\hsize}{\algocf@captiontext{#1}{#2}}% then caption is not centered
    \else%
      \global\@minipagefalse%
      \hbox to\hsize{\box\@tempboxa}% else caption is centered
    \fi%
    \addtolength{\hsize}{-\algomargin}%
  }
  \makeatother

  \jname{Biometrika}
  \markboth{N.S.~HEJAZI ET AL.}{Efficient effect estimation based on the
    generalized propensity score}

  \newcommand{\authorlist}{
      \author{N.S.~HEJAZI}
      \affil{Department of Biostatistics, T.H.~Chan School of Public Health,
      Harvard University\\
      677 Huntington Avenue, Boston, MA 02115
      \email{nhejazi@hsph.harvard.edu}}

      \author{D.C.~BENKESER}
      \affil{Department of Biostatistics \& Bioinformatics, Rollins School of
      Public Health, Emory University\\
      1518 Clifton Road, N.E.~, Atlanta, GA 30322
      \email{benkeser@emory.edu}}

      \author{I.~D\'IAZ}
      \affil{Division of Biostatistics, Department of Population Health
      Sciences, Weill Cornell Medicine\\
      402 E.~67\textsuperscript{th} Street, New York, NY 10065
      \email{ild2005@med.cornell.edu}}

      \author{M.J.~VAN DER LAAN}
      \affil{Division of Biostatistics, School of Public Health, University of
      California, Berkeley\\ 2121 Berkeley Way, Berkeley, CA 94720
      \email{laan@berkeley.edu}}
  }
}{%
  \usepackage{arxiv}
  \usepackage{standalone}
  \usepackage{multirow}
  \usepackage{setspace}
  \usepackage{amsthm}
  \usepackage[round]{natbib}

  \newtheorem{theorem}{Theorem}
  \newtheorem{lemma}{Lemma}

  {\theoremstyle{definition}\newtheorem{assumption}{}}
  {\theoremstyle{definition}\newtheorem{condition}{}}
  {\theoremstyle{definition}}

  \newcommand{\authorlist}{
    Nima S.~Hejazi \\
    Department of Biostatistics,\\
    T.H.~Chan School of Public Health,\\
    Harvard University\\
    \texttt{nhejazi@hsph.harvard.edu}\\
    \And
    David C.~Benkeser \\
    Department of Biostatistics \& Bioinformatics,\\
    Rollins School of Public Health,\\
    Emory University\\
    \texttt{benkeser@emory.edu}\\
    \And
    Iv\'an D\'iaz \\
    Division of Biostatistics,\\
    Department of Population Health Sciences,\\
    Weill Cornell Medicine\\
    \texttt{ild2005@med.cornell.edu} \\
    \And
    Mark J.~{van der Laan} \\
    Division of Biostatistics,\\
    School of Public Health,\\
    University of California, Berkeley\\
    \texttt{laan@berkeley.edu} \\
  }
}
\makeatother

\newcommand{\titlepaper}{Efficient estimation of modified treatment policy
effects based on the generalized propensity score}

\AtEndDocument{\refstepcounter{theorem}\label{finalthm}}
\AtEndDocument{\refstepcounter{equation}\label{finaleq}}
\AtEndDocument{\refstepcounter{lemma}\label{finallemma}}

\newcommand{\D}{\mathcal{D}}
\newcommand{\E}{\mathbb{E}}

\newcommand{\M}{\mathcal{M}}
\newcommand{\R}{\mathbb{R}}

\newcommand{\logit}{\text{logit}}

\newcommand{\prob}{\mathbb{P}}

\newcommand{\indep}{\mbox{$\perp\!\!\!\perp$}}
\DeclareMathOperator*{\argmin}{\arg\!\min}

\usepackage[dvipsnames]{xcolor}